\documentclass[12pt]{article}
\usepackage{bm}
\usepackage{cite}
\usepackage{mathrsfs}
\usepackage{slashed}
\usepackage{graphicx}
\usepackage{hyperref}
\usepackage{verbatim}
\usepackage{color}
\usepackage{enumerate}
\usepackage{authblk}
\usepackage[hang,flushmargin]{footmisc}
\usepackage{lipsum}
\usepackage[font=footnotesize]{caption}
\usepackage{soul}
\usepackage[utf8]{} 
\usepackage{empheq}
\usepackage{amssymb,amsmath,amsfonts}
\usepackage[makeroom]{cancel}
\usepackage[normalem]{ulem}
\usepackage{hyperref}
\usepackage{color}

\numberwithin{equation}{section}

\definecolor{blue-violet}{rgb}{0.54, 0.17, 0.89}
\definecolor{PineGreen}{cmyk}{0.92, 0, 0.59, 0.25}
\definecolor{OliveGreen}{cmyk}{0.64, 0, 0.95, 0.40}
\definecolor{RawSienna}{cmyk}{0, 0.72, 1, 0.45}
\definecolor{Gray}{cmyk}{0, 0, 0, 0.50}
\definecolor{MidnightBlue}{cmyk}{0.98, 0.13, 0, 0.43}
\definecolor{Orange}{cmyk}{0, 0.61, 0.87, 0}
\definecolor{LimeGreen}{cmyk}{0.50, 0, 1, 0}
\definecolor{Green}{cmyk}{1, 0, 1, 0}

\renewcommand{\tilde}{\widetilde}

\usepackage{mathrsfs}

\def\be{\begin{eqnarray}}
\def\ee{\end{eqnarray}}
\def\beann{\begin{eqnarray*}}
\def\eeann{\end{eqnarray*}}
\def\beq{\begin{equation}}
\def\eeq{\end{equation}}
\def\ba{\begin{array}}
\def\ea{\end{array}}
\def\ben{\begin{enumerate}}
\def\een{\end{enumerate}}
\def\bea{\begin{eqnarray}}
\def\eea{\end{eqnarray}}

\begin{document}

\title{\vspace{-70pt} \Large{\sc Non-closed scalar charge in 4-dimensional Einstein-scalar-Gauss-Bonnet black hole thermodynamics}\vspace{10pt}}
\author[a]{\normalsize{Romina Ballesteros}\footnote{\href{mailto:ext.romina.ballester@uss.cl}{ext.romina.ballester@uss.cl}}}
\author[a]{\normalsize{Marcela C\'ardenas}\footnote{\href{mailto:marcela.cardenasl@uss.cl}{marcela.cardenasl@uss.cl}}}
\author[b]{\normalsize{Eric Lescano}\footnote{\href{mailto:eric.lescano@uwr.edu.pl}{eric.lescano@uwr.edu.pl}}}

\affil[a]{\footnotesize\textit{Universidad San Sebasti\'an, Facultad de Ingenier\'ia,  Bellavista 7, Recoleta, Santiago, Chile.}}

\affil[b]{\footnotesize\textit{Institute for Theoretical Physics (IFT), University of Wroclaw, pl. Maxa Borna 9, 50-204 Wroclaw, Poland.}}

\date{}

\maketitle

\begin{abstract}
We develop a covariant differential-form framework to define scalar charges for stationary, asymptotically flat black holes in $4$--dimensional Einstein-scalar-Gauss-Bonnet
gravity with a general scalar coupling function.  Contracting the scalar field equation of motion with the horizon generator $k$ yields a non-closed-form scalar charge, revealing a bulk contribution encoded in a $3$--form, which measures the obstruction to its closedness.
In the presence of shift-symmetry, this obstruction vanishes and the $2$--form scalar
charge satisfies a Gauss law, depending solely on boundary data. Geometrically, this
reproduces known topological results in the shift-symmetric limit. This framework
allows us to analyze the role of the non-closed scalar charges in black hole thermodynamics through the Smarr formula
for more general couplings and provide a covariant, charge-based interpretation of the spontaneous scalarization mechanism, showing how the behavior of the
scalar charge and the bulk term capture the instability of scalar-free black holes
and the emergence of scalar hair. Our results offer a unified geometric understanding of the role of scalar charges and a physical interpretation to its non-closedness in terms of the spontaneous scalarization mechanism.

\end{abstract}

\newpage
\tableofcontents

\section{Introduction}
Scalar fields arise naturally in a wide range of high-energy physics contexts, particularly in the context of moduli and dilaton sectors \cite{Corvilain:2018lgw, Blumenhagen:2018nts, Rudelius:2014wla} in string theory compactifications \cite{Grana:2005jc, Blumenhagen:2006ci, Douglas:2006es}. When coupled to gravity, scalar fields typically obey no-hair theorems, which prevent them from forming non-trivial profiles around black holes unless certain conditions are considered, such as symmetry breaking \cite{Bekenstein:1974sf, Bekenstein:1975ts, Volkov:1989fi, Bizon:1990sr, Droz:1991cx, Greene:1992fw, Bekenstein:1995un,  Sotiriou:2011dz, Hui:2012qt, Babichev:2013cya, Sotiriou:2013qea, Sotiriou:2014pfa, Herdeiro:2015waa, Benkel:2016rlz}.

However, this picture changes in modified gravity theories, where scalar fields interact non-minimally with curvature invariants \cite{Kanti:1995vq, Alexeev:1996vs, Silva:2022srr}. A particular case is Einstein-scalar-Gauss-Bonnet (EsGB) gravity, where a scalar field $\phi$ couples to the Gauss-Bonnet (GB) invariant $\mathcal{G}$ through a function $f(\phi)$. This interaction arises naturally in the low-energy effective actions of string theory, where the scalar coupling often corresponds to the dilaton and the GB term captures leading $\alpha'$ corrections to the classical Einstein-Hilbert action (for a review on $\alpha'$ corrections see \cite{Lescano:2021lup}). Moreover, EsGB can be interpreted as a simple model to study the effect of a scalar field coupling to gravity.

In $4$ dimensions the GB term is a topological invariant and does not affect the dynamics unless it is coupled to a dynamical scalar. Once this coupling is introduced, it alters the dynamics and can induce scalar hair in black hole solutions, even in the absence of electric charge or angular momentum. Such black holes have been widely studied in the literature \cite{Doneva:2017bvd,Doneva:2022ewd,Silva:2017uqg,Silva:2018qhn,Julie:2019sab}. In particular, Refs.~\cite{Antoniou:2017acq,Antoniou:2017hxj} proposed a novel no-hair theorem for EsGB gravity, showing that nontrivial scalar hair can exist depending on the form of the coupling function $f(\phi)$. These models show that whether scalar hair exists or not, depends crucially on how the scalar behaves near the horizon and at spatial infinity, shaped by the choice of $f(\phi)$. 


At this point it is useful to distinguish between primary and secondary scalar hair \cite{Herdeiro:2015waa}. A primary scalar hair corresponds to an independent parameter of the black hole solution that is not determined by conserved charges or asymptotic data. By contrast, a scalar hair is referred to as 'secondary' if it depends on the conserved charges. 


In parallel to these developments, it has been shown that a covariant definition of scalar charge enables the formulation of no-hair theorems by requiring regularity both at spatial infinity and on the bifurcation surface in asymptotically flat black holes \cite{Ballesteros:2023iqb, Ballesteros:2023muf}. The key insight is to define the scalar charge through a closed $2$--form $\mathbf{Q}_\phi$. When the $4$--dimensional theory admits a shift symmetry from a real scalar,\footnote{Here by shift symmetry we mean a dependence of the action on the scalar $\phi$, such that it is invariant under $\phi+\phi_0$ where $\phi_0 $ is a constant.} one defines a current $\mathbf{J}$ associated with this symmetry and writes $d\mathbf{J}$ in terms of the scalar equation of motion $\mathbf{E}_{\phi}$, i.e., $d\mathbf{J}=-\mathbf{E}_\phi$. When evaluated on-shell\footnote{In what follows the notation $\dot{=}$ means evaluated on-shell.}  $d\mathbf{J}\dot{=}0$ and since the fields are invariant under the diffeomorphisms generated by $k$ (the Lie derivative of the current $\mathbf{J}$ with respect to $k$ vanishes $d\imath_k \mathbf{J}=0$) the closed $2$--form scalar charge is obtained through $\mathbf{Q}_\phi=\imath_k \mathbf{J}$.  

A similar procedure can be applied to theories in which the shift symmetry of a real scalar is broken by a scalar potential. In these cases, ensuring that the scalar potential is invariant under the diffeomorphisms generated by $k$ guarantees that the scalar charge is, again, closed on-shell. This construction ensures that $\mathbf{Q}_\phi$ satisfies a Gauss law: it can be integrated over any closed $2$--surface and yields the same result when evaluated either at spatial infinity or on the bifurcation surface. The definition is manifestly covariant and offers several advantages, such as allowing for a direct connection with the charges obtained in the Wald formalism and yielding a more natural appearance in the first law when considering variations of the scalar at infinity.

For generic couplings in EsGB theories, the scalar sector does not possess a shift symmetry. One might attempt to enforce the closedness of the scalar charge by requiring the Lie derivative of the coupling term $f(\phi)\mathcal{G}$ along the horizon generator to vanish. However, in stationary configurations this requirement imposes no additional constraint. The non-closedness of the $2$--form scalar charge arises because $f(\phi)$ couples directly to the Gauss–Bonnet term, rather than playing the role of a conventional scalar potential. Throughout this work, we keep $f(\phi)$ arbitrary and deal with the scalar contribution in the $2$--form scalar charge (obtained after the contraction of the scalar equation of motion with $k$) expressing it as an exact differential form plus a bulk remainder $\mathcal{W}_k$. This decomposition reveals a covariant $3$--form $\mathcal{W}_k$ that quantifies the obstruction to the closedness of the scalar charge $2$--form. We refer to the full decomposition as the non-closed-form scalar charge.

 Although for generic couplings the $2$–form scalar charge is not closed on-shell, one can still formulate no-hair statements by requiring the convergence of the corresponding surface and bulk integrals at spatial infinity. In the particular case of shift symmetry the obstruction term vanishes. Otherwise, the scalar charge ceases to be purely topological and can be interpreted using the framework of the spontaneous scalarization mechanism \cite{Doneva:2022ewd}, as we will show.  This phenomenon occurs when black hole solutions with nontrivial scalar profiles emerge dynamically, even in the absence of external sources. Indeed, it arises when a constant scalar solution becomes unstable under small scalar perturbations triggering the growth of a nontrivial scalar configuration. Spontaneous scalarization provides a mechanism for generating scalar hair without violating no-hair theorems, relying instead on a dynamical instability of the trivial scalar configuration. The process leads to genuinely non-perturbative black hole solutions and offers a framework for probing deviations from general relativity in the strong-field regime, potentially observable in gravitational wave signals \cite{Takeda:2023wqn}.

In our context, scalarization can be interpreted in terms of the asymptotic behavior of the scalar charge, through the conditions encoded in the bulk term ${\cal W}_{k}$. Indeed, it was shown that backgrounds that satisfy $\partial_\phi f(\phi_\infty)=0$ but $\partial_\phi^2f(\phi_\infty)>0$, such a process would occur. Here, we show that such particular set of linear perturbations generates a non-zero contribution to the scalar charge via ${\cal W}_{k}$, signaling the breakdown of the scalar-free configuration and the emergence of scalar hair. This provides a covariant and geometric interpretation of the mechanism, unifying it with the thermodynamic and charge-based analysis developed in this work. Indeed, our results clarify the structure of scalar charges in EsGB gravity, giving a physical interpretation to the non-clossedness of the scalar charge, and offering a systematic way to test the physical consistency of different coupling functions. They also strengthen the foundation for covariant no-hair theorems and have potential applications in string-inspired black hole models and higher-curvature gravitational theories.


We analyze the associated surface and volume integrals and prove that the scalar charge remains convergent for a wide class of coupling functions, including linear, quadratic, exponential and polynomial forms. The shift symmetric limit is defined as the case where $\mathcal{W}_k$ vanishes and the $2$--form scalar charge becomes closed on-shell satisfying a Gauss law. We also derive the Smarr formula for generic scalar couplings in the theory.


This paper is organized as follows. Section \ref{sec-theory-eom} introduces the theory and the equations of motion for the fields. In section \ref{sec-localsym} we compute the charges associated with the local symmetries (Lorentz transformations and infinitesimal diffeomorphisms) namely, the Lorentz and the Noether–Wald charges. Then, in section \ref{app-scadef} we construct the $2$--form scalar charge and show the $3$--form bulk contribution. In order to study no-hair theorems for EsGB models, in section \ref{sec-scainfybif} we study the convergence of the integrals that defines the $2$--form scalar charge at spatial infinity and over the bifurcation surface. In section \ref{Smarr} we construct the generalized Komar charge from which we can find the Smarr formula, that relates thermodynamic quantities at spatial infinity and at the event horizon for black hole spacetimes. For generic couplings the Smarr formula will have another bulk contribution and we show that in the specific dilatonic coupling that bulk term vanishes. We test our results for different couplings in section \ref{sec-test} and in section \ref{sec-spont} we discuss the implications of our findings for scalarized black holes. Finally, we present our conclusions and future directions in section \ref{sec-concl}.


  \section{The theory and its equations of motion}\label{sec-theory-eom}

Consider the EsGB action given by the integral of the $4$--form Lagrangian $\mathbf{L}$
\begin{equation}
S = \int \mathbf{L} = \frac{1}{16\pi G_{N}} \int \left[ -\star(e^a \wedge e^b) \wedge R_{ab} + \frac{1}{2} d\phi \wedge \star d\phi + \alpha'f(\phi)\, \mathcal{G} \right],
\label{main}
\end{equation}
where $\alpha'>0$ is a dimensionful constant parameter\footnote{This parameter is, in principle, generic and not necessarily related to string theory with dimensions $[L]^2$.}, $e^a$ is the vierbein, $R^{ab}$ is the $2$--form curvature of the Levi-Civita connection, and $\mathcal{G}$ is the $4$--form Gauss–Bonnet density written in differential forms as
\begin{equation}
\mathcal{G} = \frac{1}{2} \epsilon_{abcd} R^{ab} \wedge R^{cd}.
\end{equation}
Under a general variation of the fields
\beq\label{eq-deltaS}
\delta S = \int \left[\mathbf{E}_{a} \wedge \delta e^{a} + \mathbf{E}_{\phi} \delta \phi + d \mathbf{\Theta}\left(e^a,\delta e^a, \phi, \delta \phi \right) \right].
\eeq
Here $\mathbf{E}_a$ and $\mathbf{E}_\phi$ are the Einstein equations and the equation of motion for the scalar field, respectively, and $\mathbf{\Theta}$ is the presymplectic potential, which are given by,
\begin{subequations}
\begin{align}
  16\pi G_N \mathbf{E}_{a} &=\imath_a \star \left(e^b \wedge e^c\right)\wedge R_{bc} + \frac{1}{2} \left(\imath_a d\phi \wedge \star d\phi + d\phi \wedge \imath_a \star d\phi\right) - \mathcal{D} \Delta_{a}^{GB},\label{eq-eomeinst}\\  
   16\pi G_N  \mathbf{E}_\phi &= -d \star d\phi + \alpha'\partial_{\phi}  f(\phi) \mathcal{G},\label{eq-eomphi}\\
   16\pi G_N  \mathbf{\Theta} &= \left(-\star(e^a \wedge e^b) + 2 \alpha' f(\phi) \widetilde{R}^{ab}\right) \wedge \delta \omega_{ab} + \star d\phi \delta\phi + \Delta_{a}^{GB} \wedge \delta e^a. \label{eq-prep}
\end{align}
\end{subequations}
We have used the notation $\partial_\phi = \frac{\partial  }{\partial \phi}$ and $\mathcal{D}$ as the Lorentz covariant derivative, defined in the Appendix \ref{sec-apa}, Eq.~\eqref{eq:LorentzD_general}.
 Finally, the $2$--form dual curvature\footnote{In our conventions the Levi-Civita symbol is such that $\epsilon^{0123}= 1$ and  $\epsilon_{0123} =-1$.} is
\beq
\widetilde{R}^{ab} \equiv \tfrac12 \epsilon^{abcd} R_{cd}
\end{equation}
and
\begin{equation}\label{eq-deltagb}
\Delta_{d}^{GB}\equiv \alpha' \imath_a \left(d f(\phi)\right) \widetilde{R}^{a}{}_{d}.
\end{equation}
Note that these results coincide with those in Ref.~\cite{Ortin:2024emt} when the axion field vanishes.\footnote{For a recent and comprehensive self-contained derivation of Eqs.~\eqref{eq-eomeinst},~\eqref{eq-eomphi},~\eqref{eq-prep}, we refer the reader to Ref.~\cite{Ortin:2024emt}.}

\section{Local symmetries}\label{sec-localsym}

The action in Eq.~\eqref{main} is invariant under infinitesimal transformations (diffeomorphisms) and local Lorentz transformations. They act on the vierbein $e^{a}$, the scalar $\phi$, and on derived geometric quantities such as the spin connection $\omega^{ab}$ and the $2$--form curvature $R^{ab}$. Fields with Lorentz indices transform under diffeomorphisms via the Lie-Lorentz derivative (see for instance Ref.~\cite{Jacobson:2015uqa}) whereas Lorentz scalars transform with the standard Lie derivative. Details can be found in Appendix \ref{app-lielorentz}. 

We now compute both symmetries and their associated Noether charges. 

\subsection{Infinitesimal transformations}
The infinitesimal action of a diffeomorphism generated by a vector field $\xi$ on the fields is
\begin{subequations}\label{eq-diffs}
\begin{align}
\delta_{\xi} \phi &= -\imath_\xi d \phi\,, \\
\delta_{\xi} e^{a} &= -\left(\mathcal{D}\xi^a + P_{\xi}{}^{a}{}_{b} e^{b}\right)\,,\\
\delta_{\xi} \omega_{ab} &= -\left( \imath_{\xi} R_{ab} + \mathcal{D} P_{\xi}{}_{ab}\right)\,,
\end{align}
\end{subequations}
where 
\begin{equation}
    P_{\xi}{}_{ab} \equiv \nabla_{[a}\xi_{b]}.
\end{equation}
The transformation of the fields under local Lorentz transformations are given by 
\begin{subequations}\label{eq-lorentztransf}
\begin{align} 
\delta_{\sigma} \phi &= 0 \,, \\
\delta_{\sigma} e^{a} &= \sigma^{a}{}_{b} \, e^{b} \,, \label{eq-e1} \\
\delta_\sigma \omega_{ab} &= \mathcal{D} \sigma_{ab},\label{eq-spin1} \\
\delta_{\sigma} R^{ab} &= 2\sigma^{[a|}{}_{c} R^{c|b]},\label{eq-curv1}
\end{align}
\end{subequations}
where the antisymmetric parameter $\sigma_{ab} = -\sigma_{ba}$ is the Lorentz symmetry generator. The scalar field is a Lorentz scalar and remains invariant, while the vierbein, spin connection and curvature transform according to Eqs.~\eqref{eq-e1}, \eqref{eq-spin1} and \eqref{eq-curv1}. In particular, the transformation of the curvature follows from varying the spin connection in Eq.~\eqref{eq-spin1}.
\subsection{Local charges}
\subsubsection{Lorentz charge}
Considering the general variation of the fields Eq.~\eqref{eq-deltaS}, the theory is exactly invariant under local Lorentz transformations Eqs.~\eqref{eq-lorentztransf}.
It is straightforward to verify the appearance of a total derivative
$$ \delta_\sigma S = \int d \mathbf{J}[\sigma]=0,$$
leading to the $2$--form Lorentz charge 
\begin{equation} \label{eq-lorentzcharge}
\mathbf{Q}[\sigma] = \frac{1}{16\pi G_N}
\left[
- \star (e^{a} \wedge e^{b})
+ 2 \alpha' f(\phi) \, \widetilde{R}^{ab}
\right] \sigma_{ab} \,,
\end{equation}
 which is related to the off-shell invariance of the action under infinitesimal arbitrary parameters $\sigma_{ab}$. This quantity plays a central role in the Wald entropy formula, since it is obtained by evaluating the integral of this charge over a $2$--dimensional section of the horizon with $\sigma_{ab} = n_{ab}$ (the binormal to the event horizon), as noticed in Refs.~\cite{Elgood:2020mdx, Elgood:2020nls}. In Section \ref{Smarr}, we will use this charge to compute the black hole entropy explicitly. 

\subsubsection{Noether–Wald charge (diffeomorphism charge)}

Similarly to the Lorentz charge, here we consider the general variation of the fields Eq.~\eqref{eq-deltaS} and the infinitesimal transformations in Eqs.~\eqref{eq-diffs}. The EsGB theory Eq.~\eqref{main} is invariant up to a total derivative

$$\delta_\xi S = - \int d \imath_\xi \mathbf{L}.$$
After integrating by parts and using Noether identities, we arrive at the off-shell identity 
$$ d \mathbf{J}[\xi]=0.$$
\noindent
Then, there exists a $2$--form $\mathbf{Q}[\xi]$ such that $\mathbf{J}[\xi]=d\mathbf{Q}[\xi]$ which is the Noether–Wald charge
\begin{equation}\label{eq-NW}
\mathbf{Q}[\xi] = \frac{1}{16\pi G_N} \Big\{
\left[
\star (e^{a} \wedge e^{b})
- 2 \alpha' f(\phi) \, \widetilde{R}^{ab}
\right] P_{\xi\,ab}
- \Delta_{a}^{GB} \, \xi^{a} \, \big\},
\end{equation}
where $P_{\xi\,ab}$ is the Lorentz parameter induced by $\xi$ and $\Delta_a^{GB}$ encodes terms arising from the higher-derivative couplings to curvature.  
The $f(\phi) \widetilde{R}^{ab}$ term captures the Gauss–Bonnet modification, while the $\Delta_a^{GB}$ (defined in Eq.~\eqref{eq-deltagb}) term reflects the non-minimal scalar–curvature couplings. In the Wald formalism, the surface integral of $\mathbf{Q}[\xi]$ is used to define the conserved charge associated with a vector field $\xi$. When evaluated on a $2$--sphere at spatial infinity and for the timelike Killing vector of a stationary spacetime, this charge gives the mass.

In the next section we will construct the non-closed-form scalar charge for EsGB theory, which is one of the main results of this work. The computation of the Noether-Wald and Lorentz charges are relevant in the construction of the generalized Komar charge and by consequence in the Smarr formula, as we will discuss in section {\ref{Smarr}}.

\section{Scalar charge with a bulk contribution}\label{app-scadef}
We give a detailed derivation of the $2$--form scalar charge. The contraction of the equation of motion for the scalar field Eq.~\eqref{eq-eomphi} with a Killing vector $k$ reads
\begin{equation}\label{eq-app-ikE}
   16 \pi G_N ( \imath_{k} \mathbf{E}_{\phi} ) = - \imath_{k} d \star d\phi + \alpha' \partial_{\phi} f(\phi) \imath_{k} \mathcal{G}.
\end{equation}
It is assumed that all the fields (here $\varPhi$ denotes any field) are invariant under the action of a Killing vector $k$
\begin{equation}
    \delta_{k} \varPhi = 0,
\end{equation}
where by definition $\delta_k \varPhi = -\mathcal{L}_{k} \varPhi$. Thus,
\begin{equation}
    \mathcal{L}_{k} \varPhi = - \imath_{k} d \varPhi - d \imath_k \varPhi = 0.
\end{equation}
In particular, for the scalar field $\phi$, we have
\begin{equation}
    \delta_{k} \phi = -\imath_k d\phi = 0,
\end{equation}
and for $\star d\phi$, it implies\footnote{Because for a $p$--form $\omega$ and $k$ a Killing vector, holds $\mathcal{L}_k (\star \omega)=\star \mathcal{L}_k (\omega)$.}
\begin{equation}\label{eq-app1-starphi}
    \delta_{k} \star d\phi = -\imath_{k} d \star d\phi - d \imath_k \star d\phi = 0.
\end{equation}
For the Gauss-Bonnet term,
\begin{align}\label{interior-GB}
   \imath_k \mathcal{G} 
    &=2 \imath_k R^{ab} \wedge \Tilde{R}_{ab}.
\end{align}
Using eq.~\eqref{eq-app-komega} with $\xi=k$ together with the Bianchi identity $\mathcal{D}R^{ab}=0$ we obtain 
\beq
\imath_{k} \mathcal{G} = d \mathcal{X}_{k},
\label{dX}
\eeq
with the Lorentz-scalar
\begin{equation}\label{eq-lorentzsca}
\mathcal{X}_{k} = \left(- 2P_{k}^{ab} \widetilde{R}_{ab} \right).
\end{equation}
%
%
%
%
Let us observe that in order to write $$\partial_\phi f(\phi) \imath_k \mathcal{G}= \partial_{\phi} f(\phi) d \mathcal{X}_k, $$
as a total derivative, we must consider the Leibniz rule
\begin{align}
    d \left(\partial_\phi f(\phi) \mathcal{X}_k\right) = d(\partial_\phi f(\phi)) \wedge \mathcal{X}_k + \partial_\phi f(\phi) d \mathcal{X}_k.
\end{align}
Isolating the term
\beq \label{eq-app-fphig}
\partial_{\phi} f(\phi) d \mathcal{X}_k = d \left(\partial_{\phi} f(\phi) \mathcal{X}_k \right) - d(\partial_\phi f(\phi)) \wedge \mathcal{X}_k,
\eeq
and then replacing Eqs.~\eqref{eq-app1-starphi} and \eqref{eq-app-fphig} in Eq.~\eqref{eq-app-ikE}, one obtains
\begin{equation}\label{eq-isca}
  16 \pi G_N(  \imath_k \mathbf{E}_{\phi} )= d \imath_{k} \star d\phi + \alpha' d \left(\partial_\phi f(\phi) \mathcal{X}_k \right)  - \alpha' d(\partial_{\phi} f(\phi)) \wedge \mathcal{X}_k.
\end{equation}
If we name the 3-form
\beq\label{eq-Wdef}
\mathcal{W}_{k} =  d(\partial_{\phi} f(\phi)) \wedge \mathcal{X}_k,
\eeq
one can see that, in general, it is not possible to define a closed $2$-form scalar charge for EsGB theories such that $\mathcal{W}_{k} = dA_{k}$, with $A_k$ a $2$--form. However, if $\mathcal{W}_k=0$, using Eqs.~\eqref{eq-lorentzsca} and \eqref{eq-isca} on-shell, one can define the closed $2$--form scalar charge as\footnote{The normalization factor is purely conventional.}

\beq\label{eq-scadef}
 \mathbf{Q}_\phi= -\frac{1}{4\pi} \left[\imath_{k} \star d\phi - 2 \alpha' \partial_\phi f(\phi) P_{k}^{ab} \widetilde{R}_{ab}\right].
\eeq
In general, for $\mathcal{W}_k \neq 0$ it implies $d\mathbf{Q}_\phi =-\frac{\alpha'}{4\pi}\mathbf{ \mathcal{W}}_{k}$. Integrating over a $\Sigma^3$ hypersurface whose boundaries are the bifurcation surface and a $2$--sphere at spatial infinity, we apply Stokes's theorem to define the balance equation for the non-closed-form scalar charge

\begin{equation}\label{eq-scabulk}
\int_{S^{2}_{\infty}} \mathbf{Q}_\phi = \int_{\mathcal{BH}} \mathbf{Q}_\phi - \frac{\alpha'}{4\pi} \int_{\Sigma^3} \mathcal{W}_k.
\end{equation}
%
In this expression, the scalar charge is defined by the surface integral of $\mathbf{Q}_\phi$
 at spatial infinity, 
 $$\Sigma \equiv \int_{S^2_{\infty}} \mathbf{Q}_\phi.$$In the derivation of the scalar charge associated with a stationary black hole in EsGB gravity with a general scalar coupling function $f(\phi)$, the quantity $\mathbf{\mathcal{W}}_k$ emerges as a key obstruction to expressing the contraction of the scalar field equation with a Killing vector $k$ as a closed differential form. The term $\mathbf{\mathcal{W}}_k$ identically vanishes when $f(\phi)$ is linear, corresponding to the shift symmetric case studied in~\cite{Prabhu:2018aun}. In that setting, the scalar charge $q_s$ can be expressed purely as a surface integral evaluated at the horizon or at spatial infinity, leading to the topological formula
\begin{equation}
    q_s = \frac{1}{2} \alpha \kappa \, \chi(\mathcal{BH}),
\end{equation}
where $\kappa$ is the surface gravity of the black hole, $\chi(\mathcal{BH})$ is the Euler characteristic of the bifurcation surface $\mathcal{BH}$ and $\alpha$ is a coupling parameter proportional to $\alpha'$. In a more general case (non-linear coupling $f(\phi)$), $\mathbf{\mathcal{W}}_k$ does not vanish and represents a bulk contribution that obstructs the scalar charge from being expressible solely in terms of boundary data. The result in Eq.~\eqref{eq-scabulk} is also mentioned in Ref.~\cite{Prabhu:2018aun}, so this procedure makes manifest when the charge is closed (pure boundary term) and when a bulk contribution appears.

Physically, $\mathcal{W}_k$ encodes how gradients of the scalar field modulate the topological term across the spacetime, introducing nontrivial structure into the charge integral. As a consequence, the scalar charge becomes sensitive to the full spacetime configuration of the scalar and metric fields, and cannot be determined without solving the coupled field equations. Therefore, the presence of $\mathcal{W}_k$ signals the breakdown of the shift symmetry and the corresponding failure of the closedness of the scalar charge. Its vanishing is both a necessary and sufficient condition for the scalar charge to be determined entirely by topological and geometric data on the boundary.
%

\section{Scalar charge at spatial infinity and over the bifurcation surface} \label{sec-scainfybif}
In this section, we explore the consequences of the bulk obstruction in the scalar charge as Eq.~\eqref{eq-scabulk}, addressed in the particular case of a static and spherically symmetric metric
\begin{equation}\label{eq-metric}
    ds^2 = e^{A(r)} dt^2 - e^{B(r)} dr^2 - r^2 d\Omega_{(2)}^2,
\end{equation}
where $d\Omega_{(2)}^2 = d\theta^2 + \sin^2\theta\, d\varphi^2$ is the line element of the unit $2$--sphere. The behavior of the metric components and the scalar field at spatial infinity is \cite{Antoniou:2017hxj}\footnote{We thank Athanasios Bakopoulos for kindly providing us with this asymptotic expansion.}

\footnotesize
\begin{subequations}\label{BH-Solution}
\begin{align}
e^A &= 1 - \frac{2M}{r} + \frac{M \Sigma^2}{12 r^3} + \frac{24 M \Sigma \alpha'\partial_\phi f(\phi_\infty) + M^2 \Sigma^2}{6 r^4} 
\nonumber \\ & + \frac{1024M\Sigma^2 \alpha'\partial_{\phi}^2 f(\phi_\infty)+ 512 M^2 \Sigma \alpha' \partial_\phi f(\phi_\infty) - 64 \Sigma^3 \alpha' \partial_\phi f(\phi_\infty) + 96 M^3 \Sigma^2 - 3 M \Sigma^4}{325 r^5}
+ \mathcal{O}(r^{-6}),
\end{align}

\begin{align}
e^B &= 1 + \frac{2M}{r} + \frac{32 M^3 - 5M\Sigma^2}{4 r^3}  
+ \frac{768 M^4 - 208 M^2 \Sigma^2 - 384 M \Sigma \alpha'\partial_\phi f(\phi_\infty) + 3 \Sigma^4}{48 r^4} \notag \\
&\quad + \frac{97 M \Sigma^4 + 192 \Sigma^3 \alpha'\partial_\phi f(\phi_\infty)  - 1536 \Sigma^2 M \alpha' \partial_\phi^2 f(\phi_\infty) - 2464 M^3 \Sigma^2  - 6144 M^2 \Sigma \alpha'\partial_\phi f(\phi_\infty) +6144M^5}{192 r^5}  \notag \\
&\quad+ \mathcal{O}(r^{-6}), 
\end{align}

\begin{align} \label{eq-asymptsca}
\phi &= \phi_\infty + \frac{\Sigma}{r} + \frac{M \Sigma}{r^2} + \frac{32 M^2 \Sigma - \Sigma^3}{24 r^3} + \frac{12 M^3 \Sigma - 24 M^2 \alpha'\partial_\phi f(\phi_\infty) - M \Sigma^3}{6 r^4} \notag \\
&\quad + \frac{9 \Sigma^5- 928 M^2 \Sigma^3-1536 M \Sigma^2 \alpha'\partial_\phi^2 f(\phi_\infty) - 4608 \Sigma M^2 \alpha'\partial_\phi^2 f(\phi_\infty) + 6144\Sigma M^4 - 12288 M^3 \alpha'\partial_\phi f(\phi_\infty) }{1920 r^5} \notag \\
& \quad + \mathcal{O}(r^{-6}).
\end{align}
\end{subequations}

\normalsize

\noindent
In this asymptotic expansions, $M$ and $\Sigma$ are two parameters associated to the mass and the scalar charge respectively, while
$\phi_{\infty}$ is the background value of $\phi$. In what follows, we will see that in order to have a nontrivial scalar charge, $\Sigma$ should become a secondary hair.

The Killing vector that generates the horizon is $k^{t} = \delta^{\mu}_{t}$ and the momentum map in the orthonormal frame is,
\begin{align}\label{eq-P10}
    P_{k}^{10}  
                &= \frac{A'}{2} e^{\frac{(A-B)}{2}}=-P_{k}^{01}.
\end{align}
For the dual of the $2$--form curvature $\widetilde{R}_{ab}$, the only relevant component is $\widetilde{R}_{01}$ (see Appendix \ref{sec-apa}),
\begin{align}\label{eq-R01}
    \widetilde{R}_{01} 
    &= \frac{e^{-B}-1}{r^2} e^2 \wedge e^3.
\end{align}
At spatial infinity, we consider the form of the scalar field in Eq.~\eqref{eq-asymptsca},  $$\phi-\phi_\infty \sim \Sigma/r$$ as the relevant asymptotic order and then perform the Taylor expansion of $f(\phi)$ about $\phi_\infty$, which reads as
\begin{equation}
f(\phi)=\sum_{k=0}^{\infty}\frac{f^{(k)}(\phi_\infty)}{k!}\,(\phi-\phi_\infty)^k .
\end{equation}
Correspondingly, the asymptotic behavior of the first derivative of the function $f(\phi)$ with respect to the scalar field is
\begin{equation}\label{eq-asympt-paf}
\partial_\phi f(\phi)=\sum_{k=1}^{\infty}\frac{f^{(k)}(\phi_\infty)}{(k-1)!}\,(\phi-\phi_\infty)^{k-1}.
\end{equation}
Evaluating $\mathbf{Q}_\phi$ at spatial infinity with Eqs.~\eqref{eq-asymptsca} and \eqref{eq-asympt-paf} %
\begin{align}
    \int_{S^2_{\infty}} \mathbf{Q}_\phi &= -\frac{1}{4\pi} \int_{S^{2}_{\infty}} \left(  \imath_{k} \star d\phi -2\alpha' \partial_{\phi} f(\phi)P_{k}^{ab}  \widetilde{R}_{ab}\right) \\
    &= \Sigma + \mathcal{O}(r^{-3}). \label{eq:Q-infinity}
\end{align}
%
%
%

On the other hand, evaluating $\mathbf{Q}_\phi$ over the bifurcation surface where $k=0$, then $\imath_k\star d\phi=0$. Moreover, for the momentum-map Eq.~\eqref{eq-Phorizon}
\beq
P_{k}^{ab} \stackrel{\mathcal{BH}}= \kappa n^{ab}, 
\eeq
where $\kappa$ is related to the Hawking temperature \(T=\frac{\kappa}{2\pi}\) and $n^{ab}$ is normalized as $n^{ab} n_{ab} = -2$. Assuming that $\phi$ is regular and constant over bifurcation surface, then\footnote{Here, we have used the Gauss-Bonnet theorem \cite{Eguchi:1980jx} for the integral $$\int_{\mathcal{BH}} n^{ab} \tilde{R}_{ab}=4\pi \chi(\mathcal{BH}),$$ and Eq.~\eqref{eq-scahorizon} holds.}
\begin{align}
\int_{\mathcal BH}\mathbf Q_\phi
&=\frac{2\alpha'}{4\pi}\int_{\mathcal{ BH}}\partial_\phi f(\phi)\,P_k^{ab}\,\widetilde R_{ab} \notag\\
&= \frac{2\alpha'}{4\pi} \partial_{\phi} f(\phi_h) \kappa \int_{\mathcal{BH}} n^{ab} \widetilde{R}_{ab} \\
&= 2 \alpha'\kappa \partial_{\phi} f(\phi_h)   \chi(\mathcal{BH}) \label{eq-scahorizon}
\end{align}
\noindent
where $\chi(\mathcal{BH})$ is the Euler characteristic of the bifurcation surface and $\phi_h$ stands for the value of the scalar field at the event horizon. In this work, the bifurcation surface has the topology of a $2$--sphere, such that  $\chi(\mathcal{BH})=2$ and
\begin{equation}
\int_{\mathcal{BH}} \mathbf{Q}_\phi = 4 \alpha' \kappa \partial_{\phi} f(\phi_h). 
\end{equation}

Now, following Eq.~\eqref{eq-scabulk}, we need to use Eqs.~\eqref{eq-P10} and \eqref{eq-R01} together with
\beq
\partial_{\phi}^2 f(\phi) = \sum_{k=2}^\infty \frac{f^{(k)}(\phi_\infty)}{(k-2)!} \left(\phi - \phi_\infty\right)^{(k-2)},
\eeq
to compute the contribution from the 3-form $\mathcal{W}_k$. 
The analysis of the asymptotic behavior of each factor in the integrand is done considering a static, spherically symmetric black hole,
\begin{align*}
\partial_r \phi &\sim -\frac{\Sigma}{r^2} + \mathcal{O}(r^{-3}), \\
e^{(A-B)/2} &\sim 1-\frac{2M}{r} + \frac{\Sigma^2}{8r^2} + \mathcal{O}(r^{-3}),\\
A'(r) &\sim \frac{2 M}{r^2} + \mathcal{O}(r^{-3}), \\
(1-e^{-B}) &\sim \frac{2M}{r} + \mathcal{O}(r^{-2}) .
\end{align*}
Replacing the above in Eq.~\eqref{eq-Wdef} 
\begin{equation}
-\frac{\alpha'}{4\pi} \int_{\Sigma^3} \mathcal{W}_k= -\alpha'\int_{r_h}^{\infty} \partial_{\phi}^2 f(\phi_\infty)\cdot\mathcal{O}(r^{-5})   dr.
\end{equation}
The integrand in the last equation behaves as $\mathcal{O}(r^{-5})$ ensuring absolute convergence for any coupling $f(\phi)$ that is smooth near $\phi_\infty$. Consequently, the scalar charge is finite at spatial infinity even when $\mathcal{W}_k\neq0$. This finite bulk piece will reappear in the next section, where we recast the relation between quantities measured at horizon and at spatial infinity into the Smarr formula.

Finally, the balance Eq.~\eqref{eq-scabulk} for the scalar charge reads,
\begin{equation}\label{eq-sigmasurfbulk}
\Sigma = 4\alpha'\,\kappa\,\partial_\phi f(\phi_h)-\frac{\alpha'}{4\pi} \int_{\Sigma^3} \mathcal{W}_k.
\end{equation}

This result can be read as a no-(primary)hair theorem. The scalar charge is not an independent parameter, but is completely fixed by horizon data and the integral of the bulk contribution. If $\Sigma = 0$, Eq.~\eqref{eq-sigmasurfbulk} enforces
$$4\,\kappa\,\partial_\phi f(\phi_h)=\frac{1}{4\pi} \int_{\Sigma^3} \mathcal{W}_k.$$

Hence, the value of $\partial_{\phi} f(\phi_h)$ is fixed by the bulk integral and the temperature. When the bulk term vanishes, implies either $T=0$ (extremal) or $\partial_{\phi} f(\phi_h)=0$.

\section{Generalized Komar charge and Smarr formula }
\label{Smarr}
The Smarr formula was first derived as a consequence of the homogeneity of the entropy and Euler's theorem \cite{Smarr:1972kt}, relating black hole thermodynamic quantities measured at spatial infinity and at the horizon. Komar integrals \cite{Komar:1958wp} have been demonstrated to be useful for deriving the Smarr formula for stationary black holes \cite{Kastor:2010gq}. Following the framework presented in Ref.~\cite{Liberati:2015xcp}, one can construct the Komar integrals within the covariant formalism \cite{Lee:1990nz,Wald:1993nt,Iyer:1994ys}, resulting in an expression that combines surface and volume integrals. The analysis made in Ref.~\cite{Liberati:2015xcp} is applied in Lovelock theories with constant couplings for static black holes. Additionally, in Ref.~\cite{Ortin:2021ade} is shown that for Lovelock and $f(R)$ theories, the volume integral can always be rearranged as a surface term, as a consequence of a generalized Komar charge which is always closed on-shell by construction. It yields in a Smarr formula purely from surface integrals. Let us comment on the procedure: one starts computing the generalized $2$--form Komar charge, which is by definition \cite{Ortin:2021ade}
\begin{equation}
\mathbf{K}[k] \equiv - \mathbf{Q}[k] + \omega_{k},
\qquad
\imath_{k} \mathbf{L} \dot{=} d\omega_{k}.
\label{eq:Komar-def}
\end{equation}
Here $\mathbf{Q}[k]$ is the Noether-Wald charge evaluated in a Killing vector $k$ and the $2$--form $\omega_k$ is defined implicitly by taking the interior derivative with the Lagrangian evaluated on-shell. Since it is closed on-shell by construction, i.e., $d \mathbf{K}[k] = 0$, the Smarr formula is obtained integrating over a spacelike hypersurface $\Sigma^3$ whose boundaries are 
$\partial\Sigma^3 = S^2_\infty \cup \mathcal{BH}$ (a $2$--sphere at spatial infinity and the bifurcation surface). 

After using Stoke's theorem we obtain
$
\int_{S^2_\infty} \mathbf{K}[k]
= \int_{\mathcal{BH}} \mathbf{K}[k]$,
where $k$ is the Killing vector generating the horizon. In pure Einstein theory for stationary asymptotically flat spacetimes, the left-hand side is evaluated using the asymptotic form of the metric yielding the ADM mass $M$. The right-hand side is computed at the bifurcation surface, 
where $k=0$ and $P_{k\,ab} \stackrel{\mathcal{BH}}= \kappa n_{ab}$. After taking into account the generalized zeroth law \cite{Bardeen:1973gs}, the right-hand side results in two times the temperature $T$ multiplied by the Bekenstein-Hawking entropy $S$. Equating both integrals results in the usual Smarr formula $M= 2 T S$.

 Contrary to pure Einstein theory and those considered in Ref.~\cite{Ortin:2021ade}, in EsGB a bulk piece remains since the coupling $f(\phi)$ multiplies the Gauss-Bonnet invariant and it cannot be reformulated as a surface term. Indeed, the term $\imath_{k} \mathbf{L}$ evaluated on-shell is the responsible of the bulk contribution for the generalized Komar charge. By consequence, an unavoidable volume term will appear in the Smarr formula.

 In this section we will derive the Smarr formula for stationary, asymptotically flat black holes in EsGB gravity following the mentioned approaches.

We first take the trace of the Einstein Eq.~\eqref{eq-eomeinst}, isolate the Lagrangian and
evaluate it on-shell
\begin{align}
(16\pi G_N)( e^a \wedge \mathbf{E}_a) &= (16\pi G_N)(-2\mathbf{L}) + 2 \alpha' f(\phi) \mathcal{G} + d\left(e^a \wedge \Delta_a^{GB}\right) + (16\pi G_N)\mathbf{E}_{\phi} \notag \\
 &  +d\star d\phi- \alpha' \partial_{\phi} f(\phi) \mathcal{G}\\
 (16\pi G_N)\mathbf{L} &\dot{=}  \alpha' f(\phi) \mathcal{G} + \frac{1}{2}d (e^a \wedge \Delta_{a}^{GB}) + \frac{1}{2} d \star d\phi - \frac{1}{2} \alpha' \partial_\phi f(\phi) \mathcal{G}
\end{align}
and after taking the interior derivative of the Lagrangian, using the invariance of the fields under Killing vectors $k$ and the Leibniz rule, we obtain the following expression for $\imath_k 
\mathbf{L}$
\begin{align}
  (16\pi G_N)  \imath_k \mathbf{L} &\dot{=} \alpha' f(\phi) d\mathcal{X}_k - \frac{1}{2}\alpha'\partial_{\phi} f(\phi) d\mathcal{X}_k- \frac{1}{2}d \imath_k \left(e^a \wedge \Delta_a^{GB}  \right)- \frac{1}{2} d \imath_k \star d\phi\\
    &= d \Big[\alpha'\big(f(\phi) - \frac{1}{2}\partial_\phi f(\phi)\big) \mathcal{X}_k - \frac{1}{2} \imath_k \left(e^a \wedge \Delta_a^{GB}  \right)- \frac{1}{2}  \imath_k \star d\phi \Big] \notag \\
    & \hspace{1cm}+ \alpha' d \left[\frac{1}{2} \partial_\phi f(\phi) - f(\phi)\right] \wedge \mathcal{X}_k.
\end{align}
With the Noether-Wald charge in Eq.~\eqref{eq-NW} the exterior derivative of the generalized Komar charge is given by 
\begin{align}\label{eq-genkomar1}
    d\mathbf{K}[k] &= -d \mathbf{Q}[k] + \imath_k \mathbf{L} \big|_{\text{on-shell}} \\
    &=\frac{1}{16\pi G_N} \Big\{ d\Big[ - \star (e^a \wedge e^b)P_{k ab}+ \Delta_{a}^{GB}k^{a}  - \frac{\alpha' }{2}\partial_\phi f(\phi) \mathcal{X}_k - \frac{1}{2}  \imath_k \star d\phi  \notag\\
     & \hspace{1cm}+ \frac{1}{2} \big(-\Delta_a^{GB} k^{a} + e^{a} \wedge \imath_k \Delta_a^{GB} \big)\Big]+ \alpha' d \left[\frac{1}{2} \partial_\phi f(\phi) - f(\phi)\right] \wedge
     \mathcal{X}_k \Big\}.
\end{align}
If we define
\begin{equation}\label{eq-zwy}
  \mathcal{Y}_k \equiv    d f(\phi)\wedge \mathcal{X}_k, \quad \mathcal{Z}_k \equiv \frac{1}{2}\mathcal{W}_k - \mathcal{Y}_k,
\end{equation}
then Eq.~\eqref{eq-genkomar1}, provided $d \mathbf{K}[k]=0$ reads
\beq
d\tilde{\mathbf{K}}[k] + \frac{\alpha'}{16\pi G_N}\mathcal{Z}_k=0,
\eeq
where $\tilde{\mathbf{K}}[k]$ is the exact sector of \eqref{eq-genkomar1} and we call it the non-closed-form generalized Komar charge
\begin{equation}
\tilde{\mathbf{K}}[k]=\frac{1}{16\pi G_N} \Big(-\star(e^a \wedge e^b) P_{kab} + \alpha' \partial_{\phi} f(\phi) \widetilde{R}_{ab} P_{k}^{ab} + \frac{1}{2} \left( \Delta_a^{GB} k^{a} + e^a \wedge \imath_k \Delta_a^{GB} \right) - \frac{1}{2} \imath_k \star d\phi \Big).
\end{equation}
Integrating $d \mathbf{K}[k]$ over a $\Sigma^3$ hypersurface whose boundaries are the bifurcation surface and a $2$--sphere at spatial infinity and using Stokes' theorem, the balance equation for the Smarr formula with a bulk contribution reads
\beq\label{eq-genkomarbulk1}
\int_{S^{2}_{\infty}} \tilde{\mathbf{K}}[k] = \int_{\mathcal{BH}} \tilde{\mathbf{K}}[k] - \frac{\alpha'}{16\pi G_N^{(4)}}\int_{\Sigma^3} \mathbf{\mathcal{Z}}_k.
\eeq
 To evaluate the above balance equation \eqref{eq-genkomarbulk1}, we will consider an asymptotically flat fall-off, as given in Eq. \eqref{BH-Solution}, that include hairy black holes with a generic coupling $f(\phi)$ for EsGB. Indeed,
the integral of the non-closed-form generalized Komar charge at spatial infinity gives
\begin{equation}\label{eq-komarinf}  \int_{S^2_\infty}\tilde{ \mathbf{K}}[k] = \frac{M}{2} + \frac{\Sigma}{8
G_{N}},
\end{equation}
where $M$ is the ADM mass and $\Sigma$ is the scalar charge. On the other hand, the integral over the bifurcation surface gives
\begin{align} 
\int_{\mathcal{BH}} \tilde{\mathbf{K}}[k] &= \frac{1}{16\pi G_{N}}\int_{\mathcal{BH}} \Big(-\star \left( e^a\wedge e^b\right)P_{kab} + \alpha'\partial_{\phi} f(\phi) \widetilde{R}_{ab}P_k^{ab}\Big) \notag\\
&= \frac{\kappa}{2\pi}\frac{\mathcal{A}}{4 G_N}+ \frac{1}{16\pi G_N}\alpha' \partial_\phi f(\phi_h) \kappa 4\pi \chi(\mathcal{BH})\\
&= \frac{\kappa}{2\pi}\Big(\frac{\mathcal{A}}{4 G_N} + \frac{\alpha'\pi \partial_\phi f(\phi_h)}{G_N}\Big)\label{eq-komarbh},
\end{align}
with $\mathcal{A}$ denoting the area of the event horizon. The second term in the right-hand side of the last equation, corresponds to the scalar contribution shown in Eq.~\eqref{eq-scahorizon} and it  must not be understood as part of the Wald entropy.  

The Wald entropy $S$ for EdGB is given by the integral on the bifurcation surface of the Lorentz charge Eq.~\eqref{eq-lorentzcharge} for the parameter $\sigma_{ab} = n_{ab}$, yielding 
\begin{equation}\label{eq-entropyesgb}
    S = \frac{\mathcal{A}}{4G_N}  + \frac{2\pi\alpha' f(\phi_h)}{G_N}.
\end{equation}
 At the moment, the balance Eq.~\eqref{eq-genkomarbulk1} for the Smarr formula reads
\begin{equation}
    M + \frac{\Sigma}{4G_N} = 2TS + 2\alpha'\Big[\frac{\kappa}{2G_N} \big(  \partial_\phi f(\phi_h) - 2f(\phi_h)\big) - \frac{1}{16\pi G_N}\int_{\Sigma^3}\mathcal{Z}_k\Big],
\end{equation}
where the bulk integral in the balance Eq.~\eqref{eq-genkomarbulk1} turns out to be convergent for any coupling $f(\phi)$
\begin{align}
    -\frac{\alpha'}{16\pi G_N} \int_{\Sigma^3}\mathcal{Z}_k &=
    -\alpha' \int_{r_h}^{\infty} \Big( \frac{1}{2} \partial_{\phi}^2 f(\phi_\infty) - \partial_{\phi} f(\phi_\infty)\Big)\cdot \mathcal{O}(r^{-5})dr,
\end{align}
provided the smoothness of $f(\phi)$ near $\phi_\infty$ and considering that its fall-off appears at order $\mathcal{O}(r^{-5})$.

After some algebra, we may define the conjugate thermodynamical potential to $\alpha'$ in order to present the Smarr formula in a more familiar way\footnote{Since $\alpha'$ is a dimensionful parameter, it is expected to be a conjugate thermodynamical potential to $\alpha'$ in the Smarr formula and in the first law. See Refs.~\cite{Mitsios:2021zrn, Meessen:2022hcg}.}
\begin{equation}\label{eq-phialpha}
\Phi_{\alpha'} \equiv \Big( \frac{1}{16\pi G_N} \int_{\Sigma^3}\mathcal{Y}_k - \frac{\kappa f(\phi_h)}{G_N} \Big).
\end{equation}
Using the integral of the non-closed generalized Komar charge at spatial infinity Eq.~\eqref{eq-komarinf} and the integral of the non-closed generalized Komar charge over the bifurcation surface \eqref{eq-komarbh} in the balance equation for the Smarr formula Eq.~\eqref{eq-genkomarbulk1}, with the entropy definition in Eq.~\eqref{eq-entropyesgb} and the conjugate thermodynamical potential defined in Eq.~\eqref{eq-phialpha}, the Smarr formula for EsGB theories reads

\begin{equation}\label{smarr}
    M= 2TS + 2\alpha' \Phi_{\alpha'},
\end{equation}
with $M$ the ADM mass, $T$ the Hawking temperature, $S$ the entropy and $\Phi_{\alpha'}$ the conjugate thermodynamical potential which must be understood in the framework of extended thermodynamics.%

Remarkably, from Eq.~\eqref{eq-genkomarbulk1} when $\mathcal{Z}_k = 0$, i.e.,
\begin{equation}\label{differential-eq}
f(\phi)=\frac{1}{2}\partial_{\phi} f(\phi),
\end{equation}
there is no bulk contribution and $d \mathbf{K}[k] = d \tilde{\mathbf{K}}[k]=0$. This is the particular case for EdGB theories with the exponential coupling\footnote{Here the integration constant obtained from solving \eqref{differential-eq} is set to 1, as it can be reabsorbed in the coupling constant $\alpha'$ appearing in the action \eqref{main}. However, once testing particular couplings in this work, we will make explicit the constant $\mathcal{C}$.}
$$f(\phi)= e^{2\phi}.$$ 

Since the Smarr formula is the integral version of the first law, we vary the entropy Eq.~\eqref{eq-entropyesgb} with respect to $M$, $\phi_{\infty}$ and $\alpha'$ to obtain the first law for EsGB, which can be verified e.g., for the solution of Eq.~\eqref{eq-metric}, such that

\begin{equation}\label{eq-firstlaw}
\delta M = T \delta S - \frac{\Sigma}{4} \delta \phi_\infty + 2 \Phi_{\alpha'} \delta \alpha'.
\end{equation}
A similar result about the role of the background value of the scalar field $\phi_{\infty}$ in the first law was found in Refs.~\cite{Cardenas:2017chu, Astefanesei:2018vga}, for Einstein-Maxwell-dilaton theories.

%
%


\subsection{Particular cases}\label{sec-test}
Now we focus in studying some relevant cases, where we show some explicit formulas to the results coming from Eqs.~\eqref{eq-sigmasurfbulk} which is the $2$--form scalar charge with a bulk contribution, the entropy \eqref{eq-entropyesgb} and the conjugate thermodynamical potential \eqref{eq-phialpha}.
\subsubsection{Linear coupling: \texorpdfstring{$f(\phi)=\mathcal{C}\phi$, $\mathcal{C} \in \mathbb{R}$}{f(phi)=phi}}

The shift-symmetric case is the simplest coupling describing hairy black holes \cite{Liska:2023fdz,Sotiriou:2014pfa}. From the definition of the scalar charge, we will require the explicit form of the derivatives with respect to the scalar field,
$\partial_{\phi}f(\phi) = \mathcal{C}$ and $\partial_{\phi}^2f(\phi) = 0.$
Since the second derivative of the function with respect to the scalar field vanishes, the bulk term in the $2$--form scalar charge in Eq.~\eqref{eq-sigmasurfbulk} also vanishes. Then, the scalar charge simply is
\beq
\Sigma = 4\mathcal{C}\alpha' \kappa .
\eeq
From Eq.~\eqref{eq-entropyesgb} the entropy $S$ reads,
\begin{equation}
    S = \frac{\mathcal{A}}{4G_N} + \frac{2 \mathcal{C} \pi \alpha' \phi_h}{G_N},
\end{equation}
and the conjugate thermodynamical potential from Eq.~\eqref{eq-phialpha}
\begin{equation}
   \Phi_{\alpha'} = \frac{1}{16\pi G_N} \int_{\Sigma} \mathcal{Y}_k - \mathcal{C} \frac{\kappa \phi_h}{G_N}.
\end{equation}
With the previous results, the Smarr formula Eq.~\eqref{smarr} holds.
\subsubsection{Dilatonic coupling: \texorpdfstring{$f(\phi)=\mathcal{C}e^{2\phi}$, $\mathcal{C} \in \mathbb{R}$}{f(phi)=\mathcal{C}exp(2phi)}}

For EdGB theories, the existence of hairy black hole solutions was reported in \cite{Kanti:1995vq, Maeda:2009uy}. Similarly to the previous case, we will require to have at hand the first and second derivatives of the dilatonic coupling with respect to the scalar field $\partial_{\phi}f(\phi) = 2\mathcal{C}e^{2\phi}$ and $\partial_{\phi}^2f(\phi)= 4\mathcal{C}e^{2\phi}$. Since the second derivative does not vanish, the bulk term $\mathcal{W}_k$ in the $2$--form scalar charge Eq.~\eqref{eq-sigmasurfbulk} will have a contribution from a volume integral. Although there is no analytical expression, the scalar charge is given by
\beq
\Sigma = 8 \mathcal{C}\alpha' \kappa e^{2\phi_h} - \frac{\alpha'}{4\pi}\int_{\Sigma^3}\mathcal{W}_k.
\eeq
The entropy $S$ is computed with the formula in Eq.~\eqref{eq-entropyesgb} and reads
\begin{equation}
    S = \frac{\mathcal{A}}{4G_N} + \frac{2 \mathcal{C}\pi \alpha' e^{2\phi_h}}{G_N}.
\end{equation}
As we mentioned in section \ref{Smarr}, the Smarr formula for black holes in EdGB theories, is given by the integration of a closed generalized Komar charge: the bulk term $\mathcal{Z}_k$ Eq.~\eqref{eq-zwy} vanishes. The conjugate thermodynamical potential Eq.~\eqref{eq-phialpha} is 
\begin{equation}
   \Phi_{\alpha'} = \frac{1}{16\pi G_N} \int_{\Sigma^3} \mathcal{Y}_k - \frac{\kappa}{G_N} \mathcal{C}e^{2\phi_h}.
\end{equation}
Using the last equations, it can be verified that the Smarr formula Eq.~\eqref{smarr} is satisfied.

With the thermodynamics completed, we now turn to the scalarization mechanism in EsGB theories.

\section{Scalarization and general couplings}
\label{sec-spont}
In this section, we aim to interpret the scalarization mechanism in terms of the asymptotic behavior of the scalar charge, through the conditions encoded in the bulk term ${\cal W}_{k}$. Let us first briefly review the basic aspects of the process. 

In the context of EsGB theories, it is assumed the existence of a constant scalar background $ \phi = \phi_\infty$.\footnote{One may set $\phi_\infty=0$.} Allowing small perturbations around $\phi_\infty$ of the form
\begin{equation}\label{eq-pertphi}
\phi = \phi_{\infty} + \epsilon\, \delta\phi^{(1)} + \mathcal{O}(\epsilon^2), \qquad \text{for} \qquad \epsilon \ll 1.
\end{equation}
A Taylor expansion of the coupling function and its derivative yields
\begin{align}
f(\phi) &= f(\phi_{\infty}) + \epsilon \, \delta \phi^{(1)} \partial_{\phi} f(\phi_\infty) + \mathcal{O}(\epsilon^2), \\
\partial_\phi f(\phi) &= \partial_{\phi} f(\phi_{\infty}) + \epsilon \, \delta \phi^{(1)} \partial_{\phi}^2 f(\phi_{\infty}) + \mathcal{O}(\epsilon^2).
\end{align}
Plugging these expansions into the scalar equation of motion Eq.~\eqref{eq-eomphi} and keeping terms up to $  \mathcal{O}(\epsilon) $, we obtain
\begin{equation}\label{eq-pert-e}
\mathbf{E}_{\phi} = - \epsilon\, d \star d\, \delta \phi^{(1)} 
+ \partial_{\phi} f(\phi_{\infty})\, \mathcal{G}
+ \epsilon\, \delta \phi^{(1)} \partial_{\phi}^2 f(\phi_{\infty})\, \mathcal{G} + \mathcal{O}(\epsilon^2).
\end{equation}
Imposing that the constant scalar solution \( \phi = \phi_\infty \) satisfies the background equation, requires $\partial_\phi f(\phi_\infty) = 0$,
so that the unperturbed configuration solves the equation at zeroth order. At linear order in \( \epsilon \), the perturbation obeys the equation
\begin{equation}
d \star d\, \delta\phi^{(1)} + m^2_{\text{eff}}(r)\, \delta\phi^{(1)} = 0,
\end{equation}
where the $m^{2}_\text{eff}(r)$ is the effective mass squared and given by
$m^2_{\text{eff}}(r) = - \partial_{\phi}^2 f(\phi_\infty)\, \mathcal{G}(r)$.
Since $ \mathcal{G}(r) > 0 $ outside the event horizon of a  Schwarzschild spacetime, the sign of $ m^2_{\text{eff}} $ is determined by the second derivative of the coupling function. When this quantity is positive, $\partial_{\phi}^2 f(\phi_\infty) > 0$, the scalar perturbation becomes tachyonic and grows dynamically, indicating the onset of spontaneous scalarization and the formation of a nontrivial scalar configuration. A particular coupling in EsGB admits spontaneous scalarization when the following two conditions are simultaneously satisfied: 
\begin{enumerate}
\item The scalar background must solve the field equations,
\begin{equation}\label{eq-scalarized1}
f(\phi_\infty) = \text{const}, \qquad \partial_{\phi}f(\phi_\infty) = 0.
\end{equation}
\item The trivial configuration must be unstable under scalar perturbations,
\begin{equation}\label{eq-scal-second}
\partial_{\phi}^2f(\phi_\infty) > 0.
\end{equation}
\end{enumerate}

\normalsize
These criteria ensure the existence of a bifurcation point in the parameter space, from which new branches of hairy black hole solutions can emerge. Considering the conditions in Eqs.~\eqref{eq-scalarized1} and \eqref{eq-scal-second} one can determine that the linear and the dilatonic coupling are not subject to the scalarization mechanism. 

Inserting the perturbation of the scalar field Eq.~\eqref{eq-pertphi}
into the non-closed-form scalar charge Eq.~\eqref{eq-scabulk}, we may find a perturbed expression for that charge. After imposing the scalarization condition \eqref{eq-scalarized1}
we find that the scalar charge vanishes at zeroth order
$\mathbf{Q}_\phi^{(0)}=0$. 

\footnotesize
\begin{table}
\centering
\small{
\renewcommand{\arraystretch}{2.2} 
\setlength{\tabcolsep}{5.6pt}       
\begin{tabular}{|p{3.5cm}|p{7.5cm}|p{2.5cm}|}
\hline 
\textbf{Coupling} & \textbf{Scalar charge, Wald entropy and thermodynamic potential} & \textbf{Scalarization}  \\ 
\hline 
\textbf{Linear} & 
{%
\(\begin{aligned}[t]
\Sigma &=4 \mathcal{C} \alpha' \kappa  \\
S &= \frac{\mathcal{A}}{4 G_N} + \frac{2 \mathcal{C} \pi\alpha' \phi_h}{G_N} \\
\Phi_{\alpha'} &= \frac{1}{16 \pi G_N} \int_{\Sigma^3} \mathcal{Y}_k - \mathcal{C} \frac{\kappa \phi_h}{G_N} \\[3pt]
\end{aligned}\)
}
&
No \\ 
\hline
\textbf{Dilatonic} & {%
\(\begin{aligned}[t]
\Sigma &= 8 \mathcal{C}\alpha' \kappa e^{2\phi_h} - \frac{\alpha'}{4\pi} \int_{\Sigma^3} \mathcal{W}_k\\
S &=\frac{\mathcal{A}}{4 G_N} + \frac{2 \mathcal{C} \pi \alpha' e^{2 \phi_h}}{G_N} \\
\Phi_{\alpha'} &= \frac{1}{16\pi G_N} \int_{\Sigma^3} \mathcal{Y}_k - \mathcal{C} \frac{\kappa e^{2 \phi_h}}{G_N} \\[3pt]
\end{aligned}\)
} & No  \\ 
\hline 
$\mathcal{C}\frac{\phi^2}{2}(1+\beta \phi^2)$
& \(
\begin{aligned}[t]
\Sigma \;&= 4 \mathcal{C} \alpha' \kappa \big(\phi_h + 2\beta \phi_h^3\big) - \frac{\alpha'}{4\pi}\int_{\Sigma^3} \mathcal{W}_k
\\[2pt]
S \;&= \frac{\mathcal{A}}{4 G_N} + \frac{\mathcal{C} \pi \alpha' \phi_h^2 \big(1+\beta \phi_h^2\big)}{G_N} \\[2pt]
\Phi_{\alpha'} \;&= \frac{1}{16\pi G_N} \int_{\Sigma^3} \mathcal{Y}_k - \mathcal{C}\frac{\kappa \phi_h^2 \big(1+ \beta \phi_h^2\big)}{2 G_N}\\[3pt]
\end{aligned}
\)  & Yes \cite{Julie:2019sab}, for $\phi_\infty=0$
\\ 
\hline 
\textbf{Even polynomial} $f(\phi) = \mathcal{C} \phi^{2n}, n>
0$ & 
\(
\begin{aligned}[t]
\Sigma \;&=8 n\mathcal{C} \alpha' \kappa \phi_h^{2n-1} - \frac{\alpha'}{4\pi} \int_{\Sigma^3} \mathcal{W}_k \\[2pt]
S \;&= \frac{\mathcal{A}}{4 G_N} + \frac{2 \mathcal{C}\pi \alpha' \phi_h^{2n}}{G_N} \\[2pt]
\Phi_{\alpha'} \;&= \frac{1}{16\pi G_N} \int_{\Sigma^3} \mathcal{Y}_k - \mathcal{C}\frac{\kappa \phi_h^{2n}}{G_N} \\[3pt]
\end{aligned}
\)
 &
Yes, for $n=1$ and $\phi_\infty=0$ \cite{Doneva:2022ewd}
\\ 
\hline 
\textbf{Odd polynomial} $f(\phi) = \mathcal{C}\phi^{2n+1}$ & 
\(
\begin{aligned}[t]
\Sigma \;&= 4 \mathcal{C} \alpha' \kappa (2n+1) \phi_h^{2n} - \frac{\alpha'}{4\pi} \int_{\Sigma^3} \mathcal{W}_k \\[2pt]
S \;&= \frac{\mathcal{A}}{4 G_N} + \frac{2\pi \mathcal{C} \alpha' \phi_h^{2n+1}}{G_N}\\[2pt]
\Phi_{\alpha'} \;&= \frac{1}{16\pi G_N} \int_{\Sigma^3} \mathcal{Y}_k - \mathcal{C} \frac{\kappa \phi_h^{2n+1}}{G_N} \\[3pt]
\end{aligned}
\)
 & No  \\ 
\hline 
\end{tabular}
}
\caption{\small{Summary of results for differents couplings in EsGB theories.}}
\label{tab-esgb}
\end{table}
\normalsize

At first order, however, the same condition allows for a non-trivial contribution, 
leading to a linearized scalar charge of the form
\begin{equation}
 \Sigma^{(1)} = -\frac{1}{4\pi} \int_{S^2_{\infty}} \imath_k\!\star d(\delta\phi^{(1)})=
-\frac{\alpha'}{4\pi} \int_{\Sigma^3}\partial_\phi^{2}f(\phi_\infty)\,    d(\delta\phi^{(1)})\wedge\mathcal{X}_k\,,
    \label{eq-linearized_scalar_charge}
\end{equation}
where the equation in the middle comes from Eq.~\eqref{eq-scadef} and the right hand side term is the bulk contribution given by Eq.~\eqref{eq-Wdef}, without the surface integral at the bifurcation surface. 

Indeed, the volume integral in Eq.~\eqref{eq-linearized_scalar_charge} together with the condition Eq.~\eqref{eq-scal-second} implies that scalarization may occur since the perturbation $\delta\phi^{(1)}$ can generate a
non-vanishing perturbed scalar charge proportional to 
$\partial_\phi^{2}f(\phi_\infty)$. 
Additionally, the scalarization condition Eq.~\eqref{eq-scal-second} gives support to the non-closedness of the scalar charge and the bulk term provides a geometric measure of the instability.

We list our main results about scalar charges and scalarization for more general couplings in the table (\ref{tab-esgb}). 

\normalsize
\section{Conclusions}\label{sec-concl}

In this work we have presented a covariant analysis of the thermodynamic charges of stationary, asymptotically flat black holes in Einstein–scalar–Gauss–Bonnet gravity with a general scalar coupling function $f(\phi)$. 
A key element of our analysis has been the construction of the non-closed-form scalar charge $\mathbf{Q}_\phi$ in Eq.~\eqref{eq-scabulk}. Indeed, this is possible by considering the scalar field equation of motion contracted with the Killing generator of the horizon $k$. 
 In doing so, we uncovered a bulk term given by a $3$--form $\mathcal{W}_k$, which measures the non-exact part of the non-closed-form scalar charge. Then, the condition $\mathcal{W}_k=0$ emerges as the necessary and sufficient criterion for the $2$--form scalar charge to satisfy a Gauss law and be defined purely by boundary data. Physically, $\mathcal{W}_k$ encodes the coupling between gradients of the scalar field and the topological Gauss–Bonnet term. Its vanishing corresponds to the particular case of linear coupling, $f(\phi) = \phi$, in which the $2$--form scalar charge is closed and can be fully expressed in terms of the Euler characteristic of the bifurcation surface. For more general coupling functions $f(\phi)$, the obstruction term $\mathcal{W}_k$ does not vanish and depends on the full radial profile of the fields. Thus, it captures genuinely dynamical information beyond the purely topological and geometric data of the boundaries.

In evaluating the surface integrals that define the non-closed-form scalar charge, we observe that at spatial infinity, it remains finite for a broad class of coupling functions. The asymptotic expansion of the integrand at infinity shows that it decays at least as $\Sigma + \frac{\partial_{\phi} f(\phi_\infty)}{\mathcal{O}(r^{3})}$, ensuring its convergence and resulting in $\int_{S^{2}_{\infty}} \mathbf{Q}_\phi = \Sigma$. When evaluated at the bifurcation surface, the charge depends only on the horizon values of the scalar field and the surface gravity, yielding a finite result. Regarding the convergence of the bulk contribution, we find that the $3$--form $\mathcal{W}_k$ decays at infinity as $\mathcal{O}(r^{-5})$ for any smooth coupling $f(\phi)$ near $\phi_\infty$. This guarantees the volume integral in Eq.~\eqref{eq-scabulk} to be absolutely convergent, even when $\mathcal{W}_k \neq 0$. Hence, the scalar charge remains finite across the full spacetime, including configurations that explicitly break shift symmetry. From Eq.~\eqref{eq-sigmasurfbulk}, this finiteness provides an explanation for the wide variety of coupling functions $f(\phi)$ for which regular hairy black hole solutions have been found in the literature, as the decay properties of $\mathcal{W}_k$ ensure the convergence of the scalar charge irrespective of the specific form of the coupling.

It is worth noting that our construction of the non-closed-form scalar charge differs from the one presented in Ref.~\cite{Ortin:2024emt} for the particular coupling $f(\phi)=e^{-\phi}$ and in the absence of the axion field. There, the authors promote the constant coupling $\alpha'$ to a dynamical field $\ell_s(x)$ and add a $3$--form Lagrange–multiplier $C$ enforcing the constraint $d\ell_s=0$, thereby endowing the theory with global symmetries whose Noether $3$--currents are closed on-shell. Then, in a stationary background the $2$--form scalar charge is closed on-shell since it is constructed as the contraction of a Killing vector $k$ of the global Noether current, and there is no bulk contribution. By contrast, in this work we follow a procedure based on the equations of motion: contracting the scalar equation of motion with $\imath_k$ and reorganizing terms yields on-shell $d\mathbf{Q}[k]=-\tfrac{1}{4\pi}\mathcal{W}_k$,  which makes manifest the gravitational scalar contribution that appears in scalarized black hole models and reproduces Prabhu’s observation in Ref.~\cite{Prabhu:2018aun} that is, without shift symmetry, a bulk term remains.

Regarding the generalized Komar charge for EsGB theories, we observed that in general this charge will have a bulk contribution. The Smarr formula then acquires a volume integral encoded in the conjugate thermodynamic potential to $\alpha'$, consisting in the term $\mathcal{Y}_k$ proportional to the exterior derivative of the coupling function. In the particular case $f(\phi) = e^{2\phi}$, the Smarr formula will be fully defined by surface integrals.

As we pointed out, the theories under consideration depend explicitly on the dimensionful parameter $\alpha'$. Therefore, in order to write a consistent Smarr formula and first law, one has to introduce a conjugate thermodynamic potential $\Phi_{\alpha'}$, in the spirit of black hole chemistry \cite{Kastor:2010gq}. The entropy obtained from the integral of the Lorentz charge on the bifurcation surface, for the particular parameter $\sigma_{ab} =  n_{ab}$, was computed in Eq.~\eqref{eq-entropyesgb}. It is important to emphasize that the Wald entropy is related to the Lorentz charge, and not with the contribution from the Noether–Wald charge. This distinction can be easily seen from the integration of the non-closed generalized Komar charge over the bifurcation surface in Eq.~\eqref{eq-komarbh}: it yields a term proportional to the surface gravity that cannot be identified with $TS$. The Lorentz charge involves $f(\phi)$ itself, whereas the Noether–Wald charge depends on its derivative $\partial_\phi f(\phi)$ as can be seen from Eq.~\eqref{eq-NW}. The inclusion of $\Phi_{\alpha'}$ is then mandatory in order to recover the standard form of the Smarr formula in Eq.~\eqref{smarr}, with $S$ being the Wald entropy. The resulting conjugate thermodynamic potential will have the volume integral that accounts for the non-exactness of the generalized Komar charge. The same argument applies to the first law in Eq.~\eqref{eq-firstlaw}, which is obtained by varying the Wald entropy with respect to the mass, the value of the scalar at infinity and $\alpha'$.

Since our non-closed scalar charge was constructed from the equations of motion, it will have implications for the spontaneous scalarization mechanism. Within our framework, these conditions have a natural charge-based interpretation: the background scalar charge vanishes at zeroth order, but first-order perturbations generate a nontrivial scalar charge $\Sigma$. In this sense, the onset of scalarization can be viewed as the dynamical generation of scalar charge via a non-vanishing $\mathcal{W}_k$.

%

 In view of Eq.~\eqref{eq-scabulk} it is natural to ask whether a non-vanishing bulk term $\mathcal{W}_k$ may have implications for phase transitions in EsGB black holes.\footnote{ See for instance Ref.~\cite{Wu:2022xmp}, where phase transitions in asymptotically flat Lovelock black holes are shown.} For a fixed coupling function $f(\phi)$, $\mathcal{W}_k$ is a function of the fields and generally takes different values when evaluated on distinct branches of solutions within the same theory. Since it enters in the balance equation of the scalar charge between asymptotic and horizon quantities, it may affect the scalar contribution of the thermodynamic potentials in a given ensemble. A systematic comparison of thermodynamic potentials across branches would be required to determine whether the role that the bulk term plays in specific models.
Another natural question concerns the extension of the present construction to non-stationary spacetimes. In that setting, the event horizon of a dynamical black hole is no longer a Killing horizon and, in general, does not admit a bifurcation surface $\mathcal{BH}$ on which the horizon generator $k$ vanishes. As a consequence, the steps in our derivation of charges, that rely on the existence of a Killing vector $k$ and on identities evaluated at $\mathcal{BH}$, do not directly apply. In this regime, one may construct extensions of our formalism to radiative or dynamical settings (see, e.g.,~\cite{Wald:1999wa,Ashtekar:2003hk,Booth:2003ji}). A systematic treatment of this case is left for future work.

 Beside the previous extensions,\footnote{We thank our anonymous referee for bringing to our attention to these interesting relations.} other concrete generalizations of our framework deserve further investigation. It would be natural to extend the present analysis to include rotation, where additional contributions to the Smarr formula and the first law will appear and the structure of $\mathcal{W}_k$, leading to a richer scenario. Another interesting direction is to study the near-horizon geometry in the presence of $\mathcal{W}_k \neq 0$, to see whether any universal relation emerges, when the asymptotic region is modified. Finally, exploring the holographic interpretation of $\mathcal{W}_k$ in asymptotically AdS spacetimes could shed light on the dual field theory meaning of the bulk term and its thermodynamic consequences \cite{Caceres:2023gfa, Paul:2024rto}.
\newpage 
\subsection*{Acknowledgements}
 The work of RB is supported by the Vicerrectoría de Investigación y Doctorados of Universidad San Sebastián through the postdoctoral project USS-FIN-25-PDOC-05. E.L. is supported by the SONATA BIS grant 2021/42/E/ST2/00304 from the National Science Centre (NCN), Poland. RB would like to thank Felipe Díaz, Oriana Labrin, and Francisco Colipí for their helpful comments along the development of this work, as well as Matteo Zatti and Pablo Cano for enlightening discussions.

\appendix
\section{Vierbein, connection and curvature components for static, spherically symmetric metrics}\label{sec-apa}

We consider the following four-dimensional metric with signature $+ - - -$
\begin{equation}
\label{eq:metric}
    ds^2 = e^{A(r)} dt^2 - e^{B(r)} dr^2 - r^2 d\Omega_{(2)}^2,
\end{equation}
where $d\Omega_{(2)}^2 = d\theta^2 + \sin^2\theta\, d\varphi^2$ is the line element of the unit 2-sphere.

Following the conventions and formulas presented in Appendix~F.1 of Ref.~\cite{Ortin:2015hya}, the orthonormal co-frame (vierbein) adapted to the metric \eqref{eq:metric} is given by
\begin{equation}
\label{eq:vierbeins}
    e^0 = e^{A(r)/2} dt, \quad
    e^1 = e^{B(r)/2} dr, \quad
    e^2 = r d\theta, \quad
    e^3 = r \sin\theta d\varphi.
\end{equation}
From these expressions, and applying Cartan's first structure equation, the Levi-Civita spin connection $\omega^{a}{}_{b} = \omega_{\mu}{}^{a}{}_{b} d x^{\mu}
$ is defined through the relation
\begin{equation}
    \mathcal{D} e^{a} \equiv d e^a - \omega^{a}{}_{b} \wedge e^b = 0,
\end{equation}
where $\mathcal{D}$ is the exterior Lorentz-covariant derivative. We obtain the non-vanishing components of the spin connection 1-form,

\begin{align}
\label{eq:spin-connection}
    \omega^{0}{}_{1} &= -\frac{1}{2} A'(r) e^{-B(r)/2} e^0, \\
    \omega^{2}{}_{1} &=- \frac{1}{r} e^{-B(r)/2} e^2, \\
    \omega^{3}{}_{1} &=- \frac{1}{r} e^{-B(r)/2} e^3, \\
    \omega^{3}{}_{2} &= -\frac{1}{r}\cot\theta\, e^3.
\end{align}

Next, we compute the curvature 2-form $R^{a}{}_{b} = \frac{1}{2}R_{\mu\nu}{}{}^{a}{}_{b} dx^ \mu \wedge dx^{\nu}$ using Cartan's second structure equation 
\begin{equation}
    R^{a}{}_{b} \equiv d \omega^{a}{}_{b} - \omega^{a}{}_{c} \wedge \omega^{c}{}_{b},
\end{equation}
 where $d$ is the exterior derivative. The non-vanishing components of the curvature 2-form are

\begin{align}
\label{eq:curvature}
    R^{0}{}_{1} &= \left( \frac{1}{2} A'' + \frac{1}{4} A'^2 - \frac{1}{4} A' B' \right) e^{-B} e^0 \wedge e^1, \\
    R^{0}{}_{2} &=\frac{1}{2r} A' e^{-B} e^0 \wedge e^2, \\
    R^{0}{}_{3}&=\frac{1}{2r} A' e^{-B} e^0 \wedge e^3,\\
    R^{1}{}_{2} &= \left(- \frac{1}{2r} B' e^{-B}  \right) e^1 \wedge e^2, \\
    R^{1}{}_{3} &= \left(- \frac{1}{2r} B' e^{-B}  \right) e^1 \wedge e^3, \\
    R^{2}{}_{3} &= \frac{1}{r^2} \left(  e^{-B}-1 \right)  e^2 \wedge e^3.
\end{align}

The Lorentz covariant derivative whose action on a Lorentz tensor p-form with  $r$ contravariant and $s$ covariant indices is 
\begin{equation}
\mathcal{D}T^{a_1\cdots a_r}{}_{b_1\cdots b_s}
= dT^{a_1\cdots a_r}{}_{b_1\cdots b_s}
- \sum_{i=1}^{r}\omega^{a_i}{}_{c}\wedge T^{a_1\cdots c\cdots a_r}{}_{b_1\cdots b_s}
+ \sum_{j=1}^{s}\omega^{c}{}_{b_j}\wedge T^{a_1\cdots a_r}{}_{b_1\cdots c\cdots b_s}
\label{eq:LorentzD_general}
\end{equation}

%


\section{Lie-Lorentz derivative and momentum map}\label{app-lielorentz}
In this appendix we briefly review the Lie--Lorentz derivative of the vielbein and the spin connection. A detailed derivation can be found in section 2 of the Ref.~\cite{Elgood:2020svt} and references therein.

When a tensor carries Lorentz indices, its variation under infinitesimal transformations generated by a vector field $\xi$ is not simply given by minus the standard Lie
derivative, since it does not transform covariantly under local Lorentz transformations.
More explicitly, if $T$ is a Lorentz tensor
\begin{equation}
\delta_\xi T = - \mathcal{L}_\xi T,
\end{equation}
it is not Lorentz covariant.  Because of that, for any Lorentz tensor $T$ we
must write its variation with the Lie-Lorentz derivative which is defined as
\begin{equation}\label{eq-app-genvar}
\delta_\xi T = -\mathcal{L}_\xi T + \delta_{\sigma_{\xi}} T,
\end{equation}
where $\delta_{\sigma_\xi} T$ is a compensating local Lorentz
transformation which renders $\delta_\xi T$ Lorentz covariant.

For the 1-form vielbein $e^{a}$ in Eq.~\eqref{eq-app-genvar} reads
\begin{equation}
    \delta_{\xi} e^{a}{}= -\left(\mathcal{D}\xi^{a}+P_{\xi}{}^{a}{}_{b} e^b\right)
\end{equation}
where 
\begin{equation}
    P_{\xi}{}^{ab} = \nabla^{[a} \xi^{b]},
\end{equation}
is the Lorentz-momentum map. Similarly, for the 1-form spin connection in Eq.~\eqref{eq-app-genvar}
\begin{equation}\label{eq-app-komega}
    \delta_\xi \omega^{ab} = -\left(\imath_\xi R^{ab} +\mathcal{D}P_{\xi}{}^{ab }\right).
\end{equation}
In particular, when $\xi$ is a Killing vector $k$
\begin{equation}
    \delta_ke^{a} = 0, \qquad \delta_k \omega^{ab}=0.
\end{equation}
For stationary asymptotically flat black holes with bifurcate horizons, if $k$ is the Killing vector whose Killing horizon coincides with the event horizon, the Lorentz-momentum map is identified with the Killing bivector
$$P_{k}^{ab}= \nabla^{[a} k^{b]},$$
which evaluated on the bifurcation sphere $\mathcal{BH}$ results \cite{Wald:1984rg}
\begin{equation}\label{eq-Phorizon}
    P_{k}{}^{ab}\stackrel{\mathcal{BH}}{=}\kappa n^{ab},
\end{equation}
where $\kappa$ is the surface gravity related to the Hawking temperature by $T=\tfrac{\kappa}{2\pi}$ and $n^{ab}$ is the binormal to the event horizon normalized as $n^{ab} n_{ab} = -2$.




\begin{thebibliography}{99}
\bibitem{Corvilain:2018lgw}
P.~Corvilain, T.W.~Grimm and I.~Valenzuela, \emph{{The Swampland Distance Conjecture for K{\"a}hler moduli}}, \href{https://doi.org/10.1007/JHEP08(2019)075}{\emph{JHEP} {\bfseries 08} (2019) 075} [\href{https://arxiv.org/abs/1812.07548}{{\ttfamily 1812.07548}}].

\bibitem{Blumenhagen:2018nts}
R.~Blumenhagen, D.~Kl{\"a}wer, L.~Schlechter and F.~Wolf, \emph{{The Refined Swampland Distance Conjecture in Calabi-Yau Moduli Spaces}}, \href{https://doi.org/10.1007/JHEP06(2018)052}{\emph{JHEP} {\bfseries 06} (2018) 052} [\href{https://arxiv.org/abs/1803.04989}{{\ttfamily 1803.04989}}].

\bibitem{Rudelius:2014wla}
T.~Rudelius, \emph{{On the Possibility of Large Axion Moduli Spaces}}, \href{https://doi.org/10.1088/1475-7516/2015/04/049}{\emph{JCAP} {\bfseries 04} (2015) 049} [\href{https://arxiv.org/abs/1409.5793}{{\ttfamily 1409.5793}}].

\bibitem{Grana:2005jc}
M.~Grana, \emph{{Flux compactifications in string theory: A Comprehensive review}}, \href{https://doi.org/10.1016/j.physrep.2005.10.008}{\emph{Phys. Rept.} {\bfseries 423} (2006) 91} [\href{https://arxiv.org/abs/hep-th/0509003}{{\ttfamily hep-th/0509003}}].

\bibitem{Blumenhagen:2006ci}
R.~Blumenhagen, B.~Kors, D.~Lust and S.~Stieberger, \emph{{Four-dimensional String Compactifications with D-Branes, Orientifolds and Fluxes}}, \href{https://doi.org/10.1016/j.physrep.2007.04.003}{\emph{Phys. Rept.} {\bfseries 445} (2007) 1} [\href{https://arxiv.org/abs/hep-th/0610327}{{\ttfamily hep-th/0610327}}].

\bibitem{Douglas:2006es}
M.R.~Douglas and S.~Kachru, \emph{{Flux compactification}}, \href{https://doi.org/10.1103/RevModPhys.79.733}{\emph{Rev. Mod. Phys.} {\bfseries 79} (2007) 733} [\href{https://arxiv.org/abs/hep-th/0610102}{{\ttfamily hep-th/0610102}}].

\bibitem{Bekenstein:1974sf}
J.D.~Bekenstein, \emph{{Exact solutions of Einstein conformal scalar equations}}, \href{https://doi.org/10.1016/0003-4916(74)90124-9}{\emph{Annals Phys.} {\bfseries 82} (1974) 535}.

\bibitem{Bekenstein:1975ts}
J.D.~Bekenstein, \emph{{Black Holes with Scalar Charge}}, \href{https://doi.org/10.1016/0003-4916(75)90279-1}{\emph{Annals Phys.} {\bfseries 91} (1975) 75}.

\bibitem{Volkov:1989fi}
M.S.~Volkov and D.V.~Galtsov, \emph{{NonAbelian Einstein Yang-Mills black holes}}, {\emph{JETP Lett.} {\bfseries 50} (1989) 346}.

\bibitem{Bizon:1990sr}
P.~Bizon, \emph{{Colored black holes}}, \href{https://doi.org/10.1103/PhysRevLett.64.2844}{\emph{Phys. Rev. Lett.} {\bfseries 64} (1990) 2844}.

\bibitem{Droz:1991cx}
S.~Droz, M.~Heusler and N.~Straumann, \emph{{New black hole solutions with hair}}, \href{https://doi.org/10.1016/0370-2693(91)91592-J}{\emph{Phys. Lett. B} {\bfseries 268} (1991) 371}.

\bibitem{Greene:1992fw}
B.R.~Greene, S.D.~Mathur and C.M.~O'Neill, \emph{{Eluding the no hair conjecture: Black holes in spontaneously broken gauge theories}}, \href{https://doi.org/10.1103/PhysRevD.47.2242}{\emph{Phys. Rev. D} {\bfseries 47} (1993) 2242} [\href{https://arxiv.org/abs/hep-th/9211007}{{\ttfamily hep-th/9211007}}].

\bibitem{Bekenstein:1995un}
J.D.~Bekenstein, \emph{{Novel {\textquoteleft}{\textquoteleft}no-scalar-hair{\textquoteright}{\textquoteright} theorem for black holes}}, \href{https://doi.org/10.1103/PhysRevD.51.R6608}{\emph{Phys. Rev. D} {\bfseries 51} (1995) R6608}.

\bibitem{Sotiriou:2011dz}
T.P.~Sotiriou and V.~Faraoni, \emph{{Black holes in scalar-tensor gravity}}, \href{https://doi.org/10.1103/PhysRevLett.108.081103}{\emph{Phys. Rev. Lett.} {\bfseries 108} (2012) 081103} [\href{https://arxiv.org/abs/1109.6324}{{\ttfamily 1109.6324}}].

\bibitem{Hui:2012qt}
L.~Hui and A.~Nicolis, \emph{{No-Hair Theorem for the Galileon}}, \href{https://doi.org/10.1103/PhysRevLett.110.241104}{\emph{Phys. Rev. Lett.} {\bfseries 110} (2013) 241104} [\href{https://arxiv.org/abs/1202.1296}{{\ttfamily 1202.1296}}].

\bibitem{Babichev:2013cya}
E.~Babichev and C.~Charmousis, \emph{{Dressing a black hole with a time-dependent Galileon}}, \href{https://doi.org/10.1007/JHEP08(2014)106}{\emph{JHEP} {\bfseries 08} (2014) 106} [\href{https://arxiv.org/abs/1312.3204}{{\ttfamily 1312.3204}}].

\bibitem{Sotiriou:2013qea}
T.P.~Sotiriou and S.-Y.~Zhou, \emph{{Black hole hair in generalized scalar-tensor gravity}}, \href{https://doi.org/10.1103/PhysRevLett.112.251102}{\emph{Phys. Rev. Lett.} {\bfseries 112} (2014) 251102} [\href{https://arxiv.org/abs/1312.3622}{{\ttfamily 1312.3622}}].

\bibitem{Sotiriou:2014pfa}
T.P.~Sotiriou and S.-Y.~Zhou, \emph{{Black hole hair in generalized scalar-tensor gravity: An explicit example}}, \href{https://doi.org/10.1103/PhysRevD.90.124063}{\emph{Phys. Rev. D} {\bfseries 90} (2014) 124063} [\href{https://arxiv.org/abs/1408.1698}{{\ttfamily 1408.1698}}].

\bibitem{Herdeiro:2015waa}
C.A.R.~Herdeiro and E.~Radu, \emph{{Asymptotically flat black holes with scalar hair: a review}}, \href{https://doi.org/10.1142/S0218271815420146}{\emph{Int. J. Mod. Phys. D} {\bfseries 24} (2015) 1542014} [\href{https://arxiv.org/abs/1504.08209}{{\ttfamily 1504.08209}}].

\bibitem{Benkel:2016rlz}
R.~Benkel, T.P.~Sotiriou and H.~Witek, \emph{{Black hole hair formation in shift-symmetric generalised scalar-tensor gravity}}, \href{https://doi.org/10.1088/1361-6382/aa5ce7}{\emph{Class. Quant. Grav.} {\bfseries 34} (2017) 064001} [\href{https://arxiv.org/abs/1610.09168}{{\ttfamily 1610.09168}}].

\bibitem{Kanti:1995vq}
P.~Kanti, N.E.~Mavromatos, J.~Rizos, K.~Tamvakis and E.~Winstanley, \emph{{Dilatonic black holes in higher curvature string gravity}}, \href{https://doi.org/10.1103/PhysRevD.54.5049}{\emph{Phys. Rev. D} {\bfseries 54} (1996) 5049} [\href{https://arxiv.org/abs/hep-th/9511071}{{\ttfamily hep-th/9511071}}].

\bibitem{Alexeev:1996vs}
S.O.~Alexeev and M.V.~Pomazanov, \emph{{Black hole solutions with dilatonic hair in higher curvature gravity}}, \href{https://doi.org/10.1103/PhysRevD.55.2110}{\emph{Phys. Rev. D} {\bfseries 55} (1997) 2110} [\href{https://arxiv.org/abs/hep-th/9605106}{{\ttfamily hep-th/9605106}}].

\bibitem{Silva:2022srr}
H.O.~Silva, A.~Ghosh and A.~Buonanno, \emph{{Black-hole ringdown as a probe of higher-curvature gravity theories}}, \href{https://doi.org/10.1103/PhysRevD.107.044030}{\emph{Phys. Rev. D} {\bfseries 107} (2023) 044030} [\href{https://arxiv.org/abs/2205.05132}{{\ttfamily 2205.05132}}].

\bibitem{Lescano:2021lup}
E.~Lescano, \emph{{{\ensuremath{\alpha}}'-corrections and their double formulation}}, \href{https://doi.org/10.1088/1751-8121/ac463f}{\emph{J. Phys. A} {\bfseries 55} (2022) 053002} [\href{https://arxiv.org/abs/2108.12246}{{\ttfamily 2108.12246}}].

\bibitem{Doneva:2017bvd}
D.D.~Doneva and S.S.~Yazadjiev, \emph{{New Gauss-Bonnet Black Holes with Curvature-Induced Scalarization in Extended Scalar-Tensor Theories}}, \href{https://doi.org/10.1103/PhysRevLett.120.131103}{\emph{Phys. Rev. Lett.} {\bfseries 120} (2018) 131103} [\href{https://arxiv.org/abs/1711.01187}{{\ttfamily 1711.01187}}].

\bibitem{Doneva:2022ewd}
D.D.~Doneva, F.M.~Ramazano{\u{g}}lu, H.O.~Silva, T.P.~Sotiriou and S.S.~Yazadjiev, \emph{{Spontaneous scalarization}}, \href{https://doi.org/10.1103/RevModPhys.96.015004}{\emph{Rev. Mod. Phys.} {\bfseries 96} (2024) 015004} [\href{https://arxiv.org/abs/2211.01766}{{\ttfamily 2211.01766}}].

\bibitem{Silva:2017uqg}
H.O.~Silva, J.~Sakstein, L.~Gualtieri, T.P.~Sotiriou and E.~Berti, \emph{{Spontaneous scalarization of black holes and compact stars from a Gauss-Bonnet coupling}}, \href{https://doi.org/10.1103/PhysRevLett.120.131104}{\emph{Phys. Rev. Lett.} {\bfseries 120} (2018) 131104} [\href{https://arxiv.org/abs/1711.02080}{{\ttfamily 1711.02080}}].

\bibitem{Silva:2018qhn}
H.O.~Silva, C.F.B.~Macedo, T.P.~Sotiriou, L.~Gualtieri, J.~Sakstein and E.~Berti, \emph{{Stability of scalarized black hole solutions in scalar-Gauss-Bonnet gravity}}, \href{https://doi.org/10.1103/PhysRevD.99.064011}{\emph{Phys. Rev. D} {\bfseries 99} (2019) 064011} [\href{https://arxiv.org/abs/1812.05590}{{\ttfamily 1812.05590}}].

\bibitem{Julie:2019sab}
F.-L.~Juli{\'e} and E.~Berti, \emph{{Post-Newtonian dynamics and black hole thermodynamics in Einstein-scalar-Gauss-Bonnet gravity}}, \href{https://doi.org/10.1103/PhysRevD.100.104061}{\emph{Phys. Rev. D} {\bfseries 100} (2019) 104061} [\href{https://arxiv.org/abs/1909.05258}{{\ttfamily 1909.05258}}].

\bibitem{Antoniou:2017acq}
G.~Antoniou, A.~Bakopoulos and P.~Kanti, \emph{{Evasion of No-Hair Theorems and Novel Black-Hole Solutions in Gauss-Bonnet Theories}}, \href{https://doi.org/10.1103/PhysRevLett.120.131102}{\emph{Phys. Rev. Lett.} {\bfseries 120} (2018) 131102} [\href{https://arxiv.org/abs/1711.03390}{{\ttfamily 1711.03390}}].

\bibitem{Antoniou:2017hxj}
G.~Antoniou, A.~Bakopoulos and P.~Kanti, \emph{{Black-Hole Solutions with Scalar Hair in Einstein-Scalar-Gauss-Bonnet Theories}}, \href{https://doi.org/10.1103/PhysRevD.97.084037}{\emph{Phys. Rev. D} {\bfseries 97} (2018) 084037} [\href{https://arxiv.org/abs/1711.07431}{{\ttfamily 1711.07431}}].

\bibitem{Ballesteros:2023iqb}
R.~Ballesteros, C.~G\'omez-Fayr\'en, T.~Ort\'\i{}n and M.~Zatti, \emph{{On scalar charges and black hole thermodynamics}}, \href{https://doi.org/10.1007/JHEP05(2023)158}{\emph{JHEP} {\bfseries 05} (2023) 158} [\href{https://arxiv.org/abs/2302.11630}{{\ttfamily 2302.11630}}].

\bibitem{Ballesteros:2023muf}
R.~Ballesteros and T.~Ort\'\i{}n, \emph{{Hairy black holes, scalar charges and extended thermodynamics}}, \href{https://doi.org/10.1088/1361-6382/ad210a}{\emph{Class. Quant. Grav.} {\bfseries 41} (2024) 055007} [\href{https://arxiv.org/abs/2308.04994}{{\ttfamily 2308.04994}}].

\bibitem{Takeda:2023wqn}
H.~Takeda, S.~Tsujikawa and A.~Nishizawa, \emph{{Gravitational-wave constraints on scalar-tensor gravity from a neutron star and black-hole binary GW200115}}, \href{https://doi.org/10.1103/PhysRevD.109.104072}{\emph{Phys. Rev. D} {\bfseries 109} (2024) 104072} [\href{https://arxiv.org/abs/2311.09281}{{\ttfamily 2311.09281}}].

\bibitem{Ortin:2024emt}
T.~Ort{\'\i}n and M.~Zatti, \emph{{On the thermodynamics of the black holes of the Cano-Ruip{\'e}rez 4-dimensional string effective action}}, \href{https://doi.org/10.1007/JHEP06(2025)026}{\emph{JHEP} {\bfseries 06} (2025) 026} [\href{https://arxiv.org/abs/2411.10417}{{\ttfamily 2411.10417}}].

\bibitem{Jacobson:2015uqa}
T.~Jacobson and A.~Mohd, \emph{{Black hole entropy and Lorentz-diffeomorphism Noether charge}}, \href{https://doi.org/10.1103/PhysRevD.92.124010}{\emph{Phys. Rev. D} {\bfseries 92} (2015) 124010} [\href{https://arxiv.org/abs/1507.01054}{{\ttfamily 1507.01054}}].

\bibitem{Elgood:2020mdx}
Z.~Elgood, D.~Mitsios, T.~Ort\'\i{}n and D.~Pere\~n\'\i{}guez, \emph{{The first law of heterotic stringy black hole mechanics at zeroth order in \ensuremath{\alpha}'}}, \href{https://doi.org/10.1007/JHEP07(2021)007}{\emph{JHEP} {\bfseries 07} (2021) 007} [\href{https://arxiv.org/abs/2012.13323}{{\ttfamily 2012.13323}}].

\bibitem{Elgood:2020nls}
Z.~Elgood, T.~Ort\'\i{}n and D.~Pere\~n\'\i{}guez, \emph{{The first law and Wald entropy formula of heterotic stringy black holes at first order in $\alpha'$}}, \href{https://doi.org/10.1007/JHEP05(2021)110}{\emph{JHEP} {\bfseries 05} (2021) 110} [\href{https://arxiv.org/abs/2012.14892}{{\ttfamily 2012.14892}}].

\bibitem{Prabhu:2018aun}
K.~Prabhu and L.C.~Stein, \emph{{Black hole scalar charge from a topological horizon integral in Einstein-dilaton-Gauss-Bonnet gravity}}, \href{https://doi.org/10.1103/PhysRevD.98.021503}{\emph{Phys. Rev. D} {\bfseries 98} (2018) 021503} [\href{https://arxiv.org/abs/1805.02668}{{\ttfamily 1805.02668}}].

\bibitem{Eguchi:1980jx}
T.~Eguchi, P.B.~Gilkey and A.J.~Hanson, \emph{{Gravitation, Gauge Theories and Differential Geometry}}, \href{https://doi.org/10.1016/0370-1573(80)90130-1}{\emph{Phys. Rept.} {\bfseries 66} (1980) 213}.

\bibitem{Smarr:1972kt}
L.~Smarr, \emph{{Mass formula for Kerr black holes}}, \href{https://doi.org/10.1103/PhysRevLett.30.71}{\emph{Phys. Rev. Lett.} {\bfseries 30} (1973) 71}.

\bibitem{Komar:1958wp}
A.~Komar, \emph{{Covariant conservation laws in general relativity}}, \href{https://doi.org/10.1103/PhysRev.113.934}{\emph{Phys. Rev.} {\bfseries 113} (1959) 934}.

\bibitem{Kastor:2010gq}
D.~Kastor, S.~Ray and J.~Traschen, \emph{{Smarr Formula and an Extended First Law for Lovelock Gravity}}, \href{https://doi.org/10.1088/0264-9381/27/23/235014}{\emph{Class. Quant. Grav.} {\bfseries 27} (2010) 235014} [\href{https://arxiv.org/abs/1005.5053}{{\ttfamily 1005.5053}}].

\bibitem{Liberati:2015xcp}
S.~Liberati and C.~Pacilio, \emph{{Smarr Formula for Lovelock Black Holes: a Lagrangian approach}}, \href{https://doi.org/10.1103/PhysRevD.93.084044}{\emph{Phys. Rev. D} {\bfseries 93} (2016) 084044} [\href{https://arxiv.org/abs/1511.05446}{{\ttfamily 1511.05446}}].

\bibitem{Lee:1990nz}
J.~Lee and R.M.~Wald, \emph{{Local symmetries and constraints}}, \href{https://doi.org/10.1063/1.528801}{\emph{J. Math. Phys.} {\bfseries 31} (1990) 725}.

\bibitem{Wald:1993nt}
R.M.~Wald, \emph{{Black hole entropy is the Noether charge}}, \href{https://doi.org/10.1103/PhysRevD.48.R3427}{\emph{Phys. Rev. D} {\bfseries 48} (1993) R3427} [\href{https://arxiv.org/abs/gr-qc/9307038}{{\ttfamily gr-qc/9307038}}].

\bibitem{Iyer:1994ys}
V.~Iyer and R.M.~Wald, \emph{{Some properties of Noether charge and a proposal for dynamical black hole entropy}}, \href{https://doi.org/10.1103/PhysRevD.50.846}{\emph{Phys. Rev. D} {\bfseries 50} (1994) 846} [\href{https://arxiv.org/abs/gr-qc/9403028}{{\ttfamily gr-qc/9403028}}].

\bibitem{Ortin:2021ade}
T.~Ort\'\i{}n, \emph{{Komar integrals for theories of higher order in the Riemann curvature and black-hole chemistry}}, \href{https://doi.org/10.1007/JHEP08(2021)023}{\emph{JHEP} {\bfseries 08} (2021) 023} [\href{https://arxiv.org/abs/2104.10717}{{\ttfamily 2104.10717}}].

\bibitem{Bardeen:1973gs}
J.M.~Bardeen, B.~Carter and S.W.~Hawking, \emph{{The Four laws of black hole mechanics}}, \href{https://doi.org/10.1007/BF01645742}{\emph{Commun. Math. Phys.} {\bfseries 31} (1973) 161}.

\bibitem{Mitsios:2021zrn}
D.~Mitsios, T.~Ort\'\i{}n and D.~Pere\~n\'\i{}guez, \emph{{Komar integral and Smarr formula for axion-dilaton black holes versus S duality}}, \href{https://doi.org/10.1007/JHEP08(2021)019}{\emph{JHEP} {\bfseries 08} (2021) 019} [\href{https://arxiv.org/abs/2106.07495}{{\ttfamily 2106.07495}}].

\bibitem{Meessen:2022hcg}
P.~Meessen, D.~Mitsios and T.~Ort\'\i{}n, \emph{{Black hole chemistry, the cosmological constant and the embedding tensor}}, \href{https://doi.org/10.1007/JHEP12(2022)155}{\emph{JHEP} {\bfseries 12} (2022) 155} [\href{https://arxiv.org/abs/2203.13588}{{\ttfamily 2203.13588}}].

\bibitem{Cardenas:2017chu}
M.~C{\'a}rdenas, F.-L.~Juli{\'e} and N.~Deruelle, \emph{{Thermodynamics sheds light on black hole dynamics}}, \href{https://doi.org/10.1103/PhysRevD.97.124021}{\emph{Phys. Rev. D} {\bfseries 97} (2018) 124021} [\href{https://arxiv.org/abs/1712.02672}{{\ttfamily 1712.02672}}].

\bibitem{Astefanesei:2018vga}
D.~Astefanesei, R.~Ballesteros, D.~Choque and R.~Rojas, \emph{{Scalar charges and the first law of black hole thermodynamics}}, \href{https://doi.org/10.1016/j.physletb.2018.05.005}{\emph{Phys. Lett. B} {\bfseries 782} (2018) 47} [\href{https://arxiv.org/abs/1803.11317}{{\ttfamily 1803.11317}}].

\bibitem{Liska:2023fdz}
M.~Li{\v{s}}ka, R.A.~Hennigar and D.~Kubiz{\v{n}}{\'a}k, \emph{{No logarithmic corrections to entropy in shift-symmetric Gauss-Bonnet gravity}}, \href{https://doi.org/10.1007/JHEP11(2023)195}{\emph{JHEP} {\bfseries 11} (2023) 195} [\href{https://arxiv.org/abs/2309.05629}{{\ttfamily 2309.05629}}].

\bibitem{Maeda:2009uy}
K.-i.~Maeda, N.~Ohta and Y.~Sasagawa, \emph{{Black Hole Solutions in String Theory with Gauss-Bonnet Curvature Correction}}, \href{https://doi.org/10.1103/PhysRevD.80.104032}{\emph{Phys. Rev. D} {\bfseries 80} (2009) 104032} [\href{https://arxiv.org/abs/0908.4151}{{\ttfamily 0908.4151}}].

\bibitem{Wu:2022xmp}
J.~Wu and R.B.~Mann, \emph{{Thermodynamically stable phases of asymptotically flat Lovelock black holes}}, \href{https://doi.org/10.1088/1361-6382/acdd41}{\emph{Class. Quant. Grav.} {\bfseries 40} (2023) 145009} [\href{https://arxiv.org/abs/2212.08673}{{\ttfamily 2212.08673}}].

\bibitem{Wald:1999wa}
R.M.~Wald and A.~Zoupas, \emph{{A General definition of 'conserved quantities' in general relativity and other theories of gravity}}, \href{https://doi.org/10.1103/PhysRevD.61.084027}{\emph{Phys. Rev. D} {\bfseries 61} (2000) 084027} [\href{https://arxiv.org/abs/gr-qc/9911095}{{\ttfamily gr-qc/9911095}}].

\bibitem{Ashtekar:2003hk}
A.~Ashtekar and B.~Krishnan, \emph{{Dynamical horizons and their properties}}, \href{https://doi.org/10.1103/PhysRevD.68.104030}{\emph{Phys. Rev. D} {\bfseries 68} (2003) 104030} [\href{https://arxiv.org/abs/gr-qc/0308033}{{\ttfamily gr-qc/0308033}}].

\bibitem{Booth:2003ji}
I.~Booth and S.~Fairhurst, \emph{{The First law for slowly evolving horizons}}, \href{https://doi.org/10.1103/PhysRevLett.92.011102}{\emph{Phys. Rev. Lett.} {\bfseries 92} (2004) 011102} [\href{https://arxiv.org/abs/gr-qc/0307087}{{\ttfamily gr-qc/0307087}}].

\bibitem{Caceres:2023gfa}
N.~Caceres, C.~Corral, F.~Diaz and R.~Olea, \emph{{Holographic renormalization of Horndeski gravity}}, \href{https://doi.org/10.1007/JHEP05(2024)125}{\emph{JHEP} {\bfseries 05} (2024) 125} [\href{https://arxiv.org/abs/2311.04054}{{\ttfamily 2311.04054}}].

\bibitem{Paul:2024rto}
S.~Paul, S.~Gangopadhyay and A.~Saha, \emph{{Gauss{\textendash}Bonnet AdS planar and spherical black hole thermodynamics and holography}}, \href{https://doi.org/10.1088/1361-6382/ad89a0}{\emph{Class. Quant. Grav.} {\bfseries 41} (2024) 235010} [\href{https://arxiv.org/abs/2403.07543}{{\ttfamily 2403.07543}}].

\bibitem{Ortin:2015hya}
T.~Ortin, \emph{{Gravity and Strings}}, Cambridge Monographs on Mathematical Physics, Cambridge University Press, 2nd ed.~ed. (7, 2015), \href{https://doi.org/10.1017/CBO9781139019750}{10.1017/CBO9781139019750}.

\bibitem{Elgood:2020svt}
Z.~Elgood, P.~Meessen and T.~Ort\'\i{}n, \emph{{The first law of black hole mechanics in the Einstein-Maxwell theory revisited}}, \href{https://doi.org/10.1007/JHEP09(2020)026}{\emph{JHEP} {\bfseries 09} (2020) 026} [\href{https://arxiv.org/abs/2006.02792}{{\ttfamily 2006.02792}}].

\bibitem{Wald:1984rg}
R.M.~Wald, \emph{{General Relativity}}, Chicago Univ. Pr., Chicago, USA (1984), \href{https://doi.org/10.7208/chicago/9780226870373.001.0001}{10.7208/chicago/9780226870373.001.0001}.


\end{thebibliography}
\end{document}